\documentclass[aps,reprint,nofootinbib,nobibnotes,notitlepage,superscriptaddress,onecolumn,prd,
 amsmath,amssymb
]{revtex4-2}

\usepackage[caption=false]{subfig}
\usepackage{float}
\usepackage{lipsum}
\usepackage{graphicx}
\usepackage{dcolumn}
\usepackage{bm}
\usepackage{natbib}
\usepackage{hyperref}
\hypersetup{
	colorlinks = true,
    linkcolor = Red,
    urlcolor  = Red,
    citecolor = Red
}
\usepackage{booktabs}
\usepackage{braket}
\newcommand{\bk}{\mathbf{k}}
\newcommand{\bq}{\mathbf{q}}
\newcommand{\bz}{\mathbf{z}}
\newcommand{\hmpc}{h\,\mathrm{Mpc}^{-1}}

\usepackage[dvipsnames]{xcolor}

\usepackage[normalem]{ulem}

\begin{document}

\title{Searching for Folded Primordial Non-Gaussianity with Galaxy Surveys}

\author{Si-Xiang~Yang}
\email{sxyang@stanford.edu}
\affiliation{Department of Physics, Stanford University, Stanford, CA 94305, USA}
\affiliation{Kavli Institute for Particle Astrophysics and Cosmology, 382 Via Pueblo, Stanford, CA 94305, USA}

\author{Oliver~H.\,E.~Philcox}
\affiliation{Department of Physics, Stanford University, Stanford, CA 94305, USA}
\affiliation{Leinweber Institute for Theoretical Physics at Stanford, 382 Via Pueblo, Stanford, CA 94305, USA}
\affiliation{Kavli Institute for Particle Astrophysics and Cosmology, 382 Via Pueblo, Stanford, CA 94305, USA}

\begin{abstract}
\noindent Large-scale structure provides a powerful probe of inflationary physics through primordial non-Gaussianity (PNG): the galaxy power spectrum depends on local PNG through scale-dependent bias, while the galaxy bispectrum depends sensitively on both equilateral and orthogonal PNG. In this paper, we study whether galaxy surveys can also probe folded PNG, whose shape is enhanced near $k_1+k_2-k_3\rightarrow0$. We consider three inflationary models with folded PNG, including excited initial states, imaginary speeds of sound, and dissipative inflation. These models fall into two classes: cutoff-regulated cases in which the folded-enhanced region has a power-law width, and dissipation-regulated cases in which the enhanced region is exponentially narrow. We develop a numerical pipeline for computing the corresponding PNG contributions to the redshift-space galaxy power spectrum multipoles and bispectrum monopole within the EFTofLSS. Using Fisher forecasts, we show that most of the constraining power on folded PNG comes from the galaxy bispectrum. For the cutoff-regulated models, nuisance parameter marginalization causes only a mild loss of information at large folded enhancement, but finite Fourier-space binning degrades constraints once the folded region becomes narrower than the bin width. For the dissipation-regulated models, the exponentially narrow folded enhancement is hard to resolve and the observable signal instead comes from the broader support of the template, leading to weaker binning dependence but larger overlap with the equilateral template. Our results show that folded PNG is a distinctive and promising target for galaxy bispectrum analyses, and the detectability depends on the width and morphology of the folded enhancement. The numerical pipeline developed in this work is general and can be used to study a wide class of non-separable primordial bispectra.
\end{abstract}

\maketitle

\section{Introduction}
\noindent Inflation is a proposed mechanism to seed the quantum fluctuations that lead to late-time observables, such as cosmic microwave background (CMB) anisotropies and the distribution of galaxies in the large-scale structure (LSS). Those primordial relics may contain traces of physics that happened during inflation, which could have occurred at energy scales as high as $10^{15}\,\rm GeV$~\cite{Lyth:1996im,BICEP:2021xfz}, making them interesting and important cosmological probes of high-energy physics. 

In the simplest models of inflation, involving a single field with negligible self-interactions and the Bunch-Davies vacuum, the primordial curvature field obeys Gaussian statistics, and can be characterized by its two-point function only~\cite{Cabass:2016cgp, Maldacena:2002vr}
\begin{equation}
\langle\zeta(\bk_1)\zeta(\bk_2)\rangle=(2\pi)^3\delta_D(\bk_1+\bk_2) P_\zeta(k_1),
\end{equation}
where $\zeta$ is the primordial curvature field and $P_\zeta(k)$ is its primordial power spectrum. However, non-linear interactions and processes during inflation can modify the statistics of the primordial curvature field, inducing primordial non-Gaussianity (PNG) in the form of non-zero three-point functions and beyond~\cite{Chen:2006nt, Bartolo:2004if}. The three-point function of the curvature perturbation is defined as
\begin{equation}
\langle\zeta(\bk_1)\zeta(\bk_2)\zeta(\bk_3)\rangle=(2\pi)^3\delta_D(\bk_1+\bk_2+\bk_3) B_\zeta(k_1,k_2,k_3).
\end{equation}
The primordial bispectrum $B_\zeta$ can be parameterized by its amplitude $f_{\rm NL}$ and shape function $S(k_1,k_2,k_3)$: 
\begin{equation}
B_\zeta(k_1,k_2,k_3)  =\frac{18}{5}f_{\rm NL}(P_\zeta(k_1)P_\zeta(k_2)P_\zeta(k_3))^{2/3}S(k_1,k_2,k_3).
\label{eq:spf}
\end{equation}
If inflation is dilatation-invariant, the shape function depends only on the ratios $k_1/k_3$ and $k_2/k_3$, and the usual normalization scheme ensures $S=1$ for $k_1/k_3=k_2/k_3=1$ (\textit{i.e.}, $S(k,k,k)=1$). Different non-linear processes during inflation can produce different values of $f_{\rm NL}$ and shape functions of different momentum dependence. For example, having multiple fields during inflation can produce the so-called local PNG \citep[e.g.,][]{Lyth:2001nq,Enqvist:2001zp,Moroi:2001ct}, where the associated shape function is enhanced in the squeezed limit, $k_1\ll k_2\approx k_3$. For single-clock inflation, a small inflaton speed of sound can produce the equilateral and orthogonal PNG~\cite{Maldacena:2002vr,Cheung:2007st,Chen:2005fe}, where the shape function is enhanced at $k_1\approx k_2\approx k_3$ and $k_1\approx k_2\approx k_3/2$ respectively. 

Another well-motivated class of inflationary bispectra is enhanced in the folded regime, $k_1+k_2\approx k_3$~\cite{Holman:2007na, Meerburg:2009ys, Meerburg:2009fi, Aravind:2013lra, Meerburg:2015yka, Greene:2004fln, Schalm:2004qk, Chen:2010bka, Byun:2015rda, Chen:2006nt, Agarwal:2012mq, Albrecht:2014aga,Renaux-Petel:2015mga, Cremonini:2010ua, Brown:2017osf, Mizuno:2017idt, Bjorkmo:2019aev, Christodoulidis:2019mkj, Christodoulidis:2019jsx, Aragam:2019omo, Garcia-Saenz:2025jis, Aoki:2026qea, Garcia-Saenz:2018vqf, Fumagalli:2019noh, Bjorkmo:2019qno, Ferreira:2020qkf, Iarygina:2023msy,LopezNacir:2012rm,Mirbabayi:2022cbt,Green:2020whw,Salcedo:2024smn}. In the standard Bunch-Davies vacuum, curvature perturbations consist of positive-frequency mode functions only, leading to oscillatory phase factors like $e^{i(k_1+k_2+k_3)}$ in the in-in time integral, whose oscillations suppress enhancement in the folded triangles~\cite{Maldacena:2002vr,Weinberg:2005vy}. A folded enhancement can arise when the inflationary dynamics supplies a mixing of both positive- and negative-frequency components, allowing phases such as $e^{i(k_1+k_2-k_3)}$ to appear. Near $k_1+k_2-k_3\rightarrow0$, the oscillatory suppression is weakened, and the interaction can accumulate coherently over a longer conformal time interval, producing an enhanced signal along the folded regime. A familiar realization is an excited initial state, often referred to as a non-Bunch-Davies (nBD) state, in which the mode function $v_k(\eta)$ is a Bogoliubov rotation of the standard positive- and negative-frequency modes~\cite{Holman:2007na, Meerburg:2009ys, Meerburg:2009fi, Aravind:2013lra, Meerburg:2015yka, Greene:2004fln, Schalm:2004qk, Chen:2010bka, Byun:2015rda, Chen:2006nt, Agarwal:2012mq, Albrecht:2014aga},
\begin{equation}
    v_k(\eta)=\alpha_k u^+_k(\eta)+\beta_k u_k^-(\eta),
    \label{eq:intro_modefunction}
\end{equation}
where $u_k^+\propto e^{-ik\eta}$ is the positive-frequency mode function and $u_k^-=(u_k^+)^*$. The Bunch-Davies vacuum corresponds to $\beta_k=0$, while an excited initial state has $\beta_k\neq0$.
The same interference perspective also explains the folded enhancement in imaginary speed-of-sound models, which can appear as an EFT description of strongly non-geodesic motion during inflation~\cite{Renaux-Petel:2015mga, Cremonini:2010ua, Brown:2017osf, Mizuno:2017idt, Bjorkmo:2019aev, Christodoulidis:2019mkj, Christodoulidis:2019jsx, Aragam:2019omo, Garcia-Saenz:2025jis, Aoki:2026qea,Garcia-Saenz:2018vqf, Fumagalli:2019noh, Bjorkmo:2019qno, Ferreira:2020qkf, Iarygina:2023msy}. Additionally, folded PNG can arise in warm inflation~\cite{Berera:1995ie,Berera:1995wh,Berera:1996nv,Berera:1999ws,LopezNacir:2011kk,LopezNacir:2012rm,Bastero-Gil:2011rva,Bastero-Gil:2014raa,Berghaus:2019whh,Mirbabayi:2022cbt}, representing a generic feature of classically-sourced inflationary correlators~\cite{Green:2020whw, Jiang:2015hfa}. Recently, it has been shown under the open effective field theory of inflation framework that a characteristic feature of the primordial bispectrum in the weak-dissipation regime is an enhanced but finite folded limit~\cite{Salcedo:2024smn}. 

Probing PNG with galaxy clustering requires modeling and separating out non-linear gravity and galaxy clustering physics. The effective field theory of large-scale structure (EFTofLSS) offers a perturbative framework for modeling galaxy clustering on mildly non-linear scales~\cite{Baumann:2010tm, Carrasco:2012cv}. Its reach has been substantially extended through higher-loop calculations~\cite{Foreman:2015lca,Carrasco:2013mua, Baldauf:2015zga,DAmico:2022ukl,Philcox:2022frc} and higher-point correlation functions~\cite{Steele:2021lnz,Bertolini:2016bmt}, and it has been used extensively in analyzing redshift-space galaxy clustering data~\cite{Ivanov:2019pdj,DAmico:2019fhj,Chen:2021wdi,DESI:2024hhd,Chudaykin:2025aux,Bakx:2025pop,Cabass:2024wob}. One important advantage of this framework is that the effect of primordial non-Gaussianity on galaxy clustering can be systematically included, allowing the corresponding redshift-space spectra to be calculated under perturbative control. Extensions of EFTofLSS in the presence of PNG have been explored in various cases, including local, equilateral, and orthogonal shapes~\cite{Assassi:2015fma, Assassi:2015jqa}. Those PNG types have been well studied and constrained using CMB anisotropies~\cite{Planck:2019kim,Jung:2025nss} and large-scale structure survey data~\cite{Cabass:2022ymb,Cabass:2022wjy,Chudaykin:2025vdh,Ivanov:2024hgq,Chen:2024bdg,DAmico:2022gki,Chaussidon:2024qni}. Although folded PNG serves as the signal of various well-motivated inflationary physics scenarios, previous studies (with the exception of~\citep{Salcedo:2026sdn}) on folded PNG constraints have mainly relied on separable templates and projections on the standard shape templates~\cite{Planck:2019kim,Meerburg:2009fi,Meerburg:2015yka}. A consistent EFTofLSS treatment of galaxy clustering in the presence of folded PNG remains to be explored.

In this work, we study the prospects for constraining folded PNG using galaxy clustering. We consider three representative inflationary scenarios that span two classes of morphology: non-Bunch-Davies initial states and imaginary speed of sound models, in which the saturated folded-enhanced region has a power-law width, and dissipative inflation, in which the enhanced region is exponentially narrow. We develop a numerical pipeline for computing contributions from general PNG bispectra to the one-loop galaxy power-spectrum multipoles and tree-level bispectrum monopole within the EFTofLSS framework, focusing in particular on the scale-dependent bias contributions. We then utilize the pipeline for folded PNG and perform Fisher forecasts on the amplitude $f_{\rm NL}$, and investigate how nuisance-parameter marginalization and finite Fourier-space binning affect the resulting constraining power.

The remainder of this paper is structured as follows: in Sec.~\ref{sec2}, we introduce the three representative folded PNG models, organize them into cutoff-regulated and dissipation-regulated classes, and compare their characteristic features in the squeezed and folded regime. In Sec.~\ref{sec3}, we outline the EFTofLSS framework for computing the galaxy power spectrum and bispectrum in the presence of PNG. In Sec.~\ref{sec4}, we present the Fisher forecast results and discuss how the folded features affect nuisance-parameter marginalization and Fourier-space binning. Finally, we conclude in Sec.~\ref{sec5}, with further details of the theoretical model presented in Appendix~\ref{app:MD}.

\section{Inflation Models with Folded PNG Signatures}
\label{sec2}
\noindent In this section, we review three representative inflationary models with enhancements in the folded regime. In general, we start from the inflationary action, discuss the correlation functions computed using the in-in formalism, and analyze the behavior of three-point functions in the folded and squeezed regimes.  

The standard quadratic action for a scalar field minimally coupled to gravity can be written as 
\begin{equation}
    S_2=\int dtd^3x\, \frac{a^3\epsilon}{c_s^2}\left[\dot{\zeta}^2-\frac{c_s^2}{a^2}(\partial_i\zeta)^2\right],
\end{equation}
where $\epsilon=-\dot{H}/H^2$ is the slow-roll parameter and $c_s$ is the speed of sound of the curvature field. The action can be derived using the effective field theory of inflation (EFTofI)~\cite{Creminelli:2006xe, Cheung:2007st}, where a Nambu-Goldstone boson $\pi\approx-\zeta/H$ associated with the spontaneous symmetry breaking of time-translation symmetry during inflation is introduced. 
The curvature perturbation after canonical quantization becomes a superposition of creation and annihilation operators, $\sqrt{2a^2\epsilon}\,\zeta(\bk,\eta)=v_k(\eta)a_\bk+v_k^*(\eta)a_{-\bk}^\dagger$, and the most general solution to the equation of motion of $\zeta$ is a superposition of positive- and negative-frequency mode functions, as in Eq.~\eqref{eq:intro_modefunction}:
\begin{equation}
    v_k(\eta)=\alpha_ku^+_k(\eta)+\beta_ku_k^-(\eta),\quad u^+_k(\eta)=\frac{e^{-ic_sk\eta}}{\sqrt{2c_sk}}\left(1-\frac{i}{c_sk\eta}\right),\quad u_k^-=(u_k^+)^*.
    \label{eq:MSsol}
\end{equation}
It is usually assumed that the initial condition of the mode function asymptotes to the Minkowski one in the far past (where modes are deeply subhorizon), giving $|\alpha_k|=1, \beta_k=0$. The state $\ket{0}$ specified by $a_\bk\ket{0}=0$ for all $\bk$ is called the Bunch-Davies vacuum, which is the lowest-energy state of de Sitter background.

\subsection{Cutoff-Regulated Folded Enhancement}

\noindent The non-Bunch-Davies and imaginary speed-of-sound scenarios have different microscopic origins, but their folded enhancements share a common structure. In both cases, a finite initial time regulates the conformal time integral appearing in the in-in calculation of the three-point function and produces folded enhancements in the shape function, whose width decreases as a power law. We therefore discuss these models together as examples of cutoff-regulated folded enhancement. 

\subsubsection{Non-Bunch-Davies Initial States}
\noindent In a number of non-standard inflationary models, the Bunch-Davies vacuum is not the most natural choice of initial conditions. For example, if inflation is so short that it starts right before the largest observable scales exit the horizon (which could be suggested by large-scale CMB anomalies \citep[e.g.,][]{Contaldi:2003zv}), the initial conditions for these large-scale modes may deviate from Minkowski spacetime. Meanwhile, if inflation is sufficiently long, the physical wavenumber of the curvature perturbations at the start of inflation will exceed some UV cutoff scale, such as $M_\text{pl}$, and new physics effects, such as quantum gravity, could excite negative frequency modes, \textit{i.e.} $\beta_k\neq0$ in Eq.~\eqref{eq:MSsol}, resulting in a non-Bunch-Davies vacuum.

Inflationary scenarios with a non-Bunch-Davies vacuum have distinct non-Gaussianity features, peaking in the folded regime~\cite{Holman:2007na, Meerburg:2009ys, Meerburg:2009fi, Aravind:2013lra, Meerburg:2015yka, Greene:2004fln, Schalm:2004qk, Chen:2010bka, Byun:2015rda, Chen:2006nt, Agarwal:2012mq, Albrecht:2014aga}. The physical intuition is that the perturbation modes that are initially not in their lowest energy state are unstable and can decay into
longer wavelength modes, resulting in physical poles in the folded limit of $n$-point functions~\cite{Jiang:2015hfa, Green:2020whw}. There have been extensive efforts to explore different realizations of the non-Bunch-Davies vacuum (for a review of the different realizations, see~\cite{Meerburg:2015yka}). In the following, we choose one specific realization of the non-Bunch-Davies vacuum with a small speed of sound for illustration, which has a well-behaved squeezed limit and notable oscillatory behavior~\cite{Meerburg:2009fi}.

The model introduces the non-Bunch-Davies vacuum by the New Physics Hypersphere (NPH) approach~\cite{Greene:2004fln, Meerburg:2009fi}: the effective description of the full theory is assumed to be valid for $\eta$ larger than an initial time $\eta_0$, which is identified as when the physical momentum equals the
high energy cutoff $\Lambda_c\ (\equiv \lambda H)$, \textit{i.e.} $kc_s/a = \Lambda_c$. This sources a scale-dependent initial time, $\eta_0=\eta_0(k) = -\lambda/(kc_s)$ (assuming a de Sitter background), which preserves scale-invariance in the output correlators.\footnote{Another approach to introduce the non-Bunch-Davies vacuum is the boundary EFT,\cite{Schalm:2004qk}, where the initial time is fixed rather than the energy scale. This scenario is heavily constrained by the scale-invariance of the power spectrum, however.} At $\eta_0(k)$, the effects of the UV theory amount to effectively generating excited states. Assuming that the high energy effects should be suppressed by $H/\Lambda_c=1/\lambda \ll 1$, the absolute value of the Bogoliubov parameters can be written in perturbative form:
\begin{equation}
    |\alpha_k|=1+y\frac{H}{\Lambda_c}+\mathcal{O}\left(\frac{H^2}{\Lambda_c^2}\right),\quad |\beta_k|=x\frac{H}{\Lambda_c}+\mathcal{O}\left(\frac{H^2}{\Lambda_c^2}\right),
\end{equation}
where $x$ and $y$ are order-one parameters and we assume that the Bogoliubov parameters are scale-invariant. Compared to $\beta_k=0$ in the Bunch-Davies case, $v_k$ in Eq.~\eqref{eq:MSsol} now has a negative frequency component. 

The mixing of the positive- and negative-frequency mode functions results in an enhanced signal in the folded limit. Specifically, using the in-in formalism~\cite{Maldacena:2002vr, Weinberg:2005vy,Chen:2017ryl}, the tree-level three-point function can be computed as
\begin{equation}
    \langle \zeta(\bk_1,\eta)\zeta(\bk_2,\eta)\zeta(\bk_3,\eta)\rangle=2\,\text{Im}\left[\int_{\eta_0}^\eta d\eta'\langle0|\zeta(\bk_1,\eta')\zeta(\bk_2,\eta')\zeta(\bk_3,\eta')H_{I}(\eta')|0\rangle\right],
    \label{eq:inin3pt}
\end{equation}
where $H_{I}$ is the interaction picture cubic Hamiltonian. In the standard calculation, the initial time $\eta_0$ is extended to $-\infty$, with an $i\epsilon$ prescription to ensure convergence. In the case with a non-Bunch-Davies vacuum, however, the effective description of the theory is only valid after the NPH crossing time. In the following, we choose $\eta_0$ as $\eta_0=-\lambda/(k_tc_s)$, where $k_t=\sum_{i=1,2,3}k_i$.\footnote{Other choices for $\eta_0$ are valid and adopted in the literature, such as $\max(k_1,k_2,k_3)c_s|\eta_0|=\lambda$ \cite{Garcia-Saenz:2018vqf}. Different prescriptions can change the detailed phase of oscillations but do not change the scaling behavior in the folded regime.} 

The correlation function can be found by inserting $\zeta(\bk)$ and time-derivatives of $\zeta(\bk)$ into Eq.~\eqref{eq:inin3pt} and performing the conformal time integral. In the standard Bunch-Davies vacuum case, the curvature perturbation consists of positive-frequency mode functions only, which would result in an oscillatory phase as $e^{i(k_1+k_2+k_3)}$. For the non-Bunch-Davies case, however, the curvature perturbation is a superposition of positive- and negative-frequency mode functions (Eq.~\eqref{eq:MSsol}), and we can have insertions of negative-frequency mode functions. The leading contribution comes from the operator $\dot\zeta^3$ in the interaction Hamiltonian~\cite{Meerburg:2009fi}, where one insertion of the negative-frequency mode function contributes at order $\beta_k$ as 
\begin{equation}
    2\text{Im}\left[\int_{\eta_0}^\eta d\eta'\langle0|\zeta(\bk_1,\eta')\zeta(\bk_2,\eta')\zeta(\bk_3,\eta')H_{I}(\eta')|0\rangle\right]\sim|\beta_k|\sum_{j=1,2,3}\int_{\eta_0}^{\eta}d\eta'\eta'^2 e^{i\tilde{k}_jc_s\eta'},
    \label{eq:nBDint}
\end{equation}
where $\tilde{k}_j\equiv\sum_ik_i-2k_j$ appears in the exponent as a result of two positive frequency and one negative frequency mode functions. More insertions of negative-frequency mode functions are suppressed by higher powers of $\beta_k$. Working at order $\beta_k$, the resulting shape function is~\cite{Meerburg:2009fi}:
\begin{equation}
    S_{\rm nBD}(k_1,k_2,k_3)=\frac{1}{\mathcal{N}_{\rm nBD}}k_1k_2k_3\sum_{i=1,2,3}{\frac{f\left(\lambda\frac{\tilde{k}_i}{k_t}\right)}{\tilde{k}_i^3}}
    \label{eq:shapefunc}
\end{equation}
where
\begin{equation}
    f(x)=\cos(\delta)-\cos(x+\delta)+\frac{x^2\cos(x+\delta)}{2}-x\sin{(x+\delta)},\qquad\mathcal{N}_{\rm nBD}=\frac{\lambda^3\sin\delta}{192},
    \label{eq:normfactor}
\end{equation}
and $\delta$ is the phase of the Bogoliubov parameter.\footnote{The effect of the phase $\delta$ on the shape function is discussed in detail in~\cite{Meerburg:2009fi}. For the purpose of this paper, we use $\delta=\pi/3$ as a representative phase for plots and analysis, avoiding values like $\delta=\pi/2$ and $3\pi/2$, where sub-leading terms need to be included~\cite{Meerburg:2009fi}. } The normalization factor $\mathcal{N}_{\rm nBD}$ is defined such that $|S_{\rm nBD}(k/2\,,k/2\,,k)|=1$, \textit{i.e.}, the shape function is $\mathcal{O}(1)$ in the folded regime. Note that the normalization scheme adopted here is different from the usual scheme of a unit equilateral limit $S(k,k,k)=1$, and is chosen to simplify comparisons between models.

\begin{figure}[!tp]
    \centering
    \includegraphics[width=\linewidth]{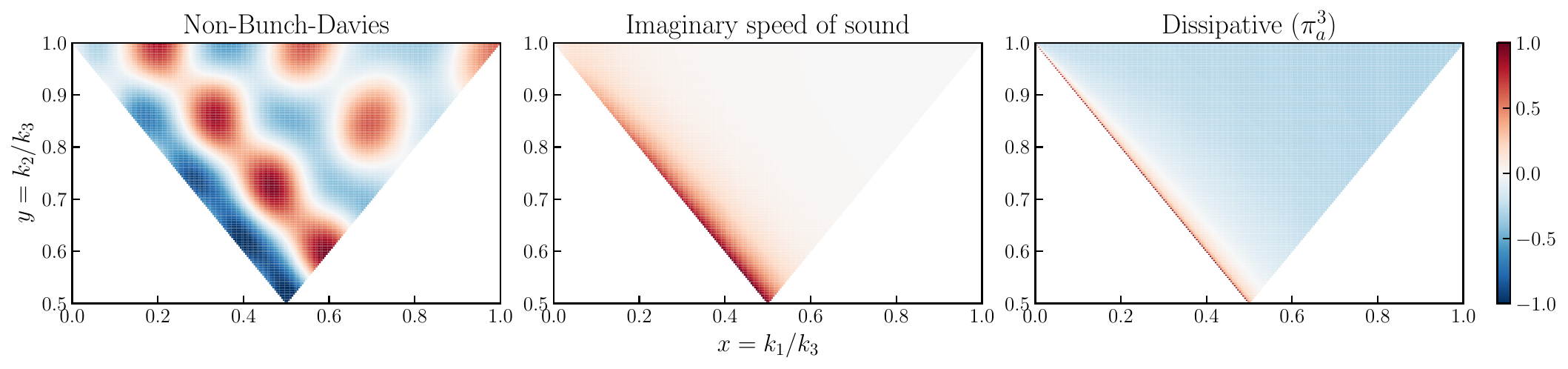}
    \caption{Shape function $S(x=k_1/k_3,y=k_2/k_3,1)$ for three scenarios: non-Bunch-Davies vacuum ($\lambda=50$),  imaginary speed of sound ($\lambda=50$), and dissipative inflation ($\pi_a^3$ operator, $H/\gamma=50$). $x=y=1$ is the equilateral limit, $x\ll y=1$ is the squeezed limit, and $x+y=1$ is the exact folded limit. The shape functions are normalized such that $|S(k/2\,,k/2\,,k)|=1$, \textit{i.e.}, the shape function is $\mathcal{O}(1)$ in the folded regime. Though all three templates demonstrate a folded enhancement, the details of the shape are different: the non-Bunch-Davies case exhibits oscillatory features as a result of the interference between positive- and negative-frequency mode functions, while dissipative inflation has a much narrower folded core, as discussed in Eq.~\eqref{eq:dissiwidth}.}
    \label{fig:EIS}
\end{figure}

The shape function is plotted in Fig.~\ref{fig:EIS}, together with the other two folded-enhanced templates discussed later. The parameter $\lambda$, defined as $\Lambda_c/H$ in this model, enters the shape function through the cutoff $\eta_0=-\lambda/(k_tc_s)$ in Eq.~\eqref{eq:nBDint} and controls the qualitative behavior of the template, including the oscillatory feature and the folded enhancement. In the following, we analyze the shape function in two important limits (the relevant behavior is summarized in Table~\ref{table:folded}):
\begin{itemize}
        \item \textbf{Folded}: The most characteristic feature of the template is that it is enhanced in the folded regime. For example, in the limit $\tilde k_3=(k_1+k_2)-k_3\rightarrow0$, the dominant contribution to the normalized shape function is
        \begin{equation}
        S_{\rm nBD}(k_1,k_2,k_3)\supset\frac{1}{\mathcal{N}_{\rm nBD}}\frac{k_1k_2k_3}{\tilde{k}_3^3}\,f\left(\lambda\frac{\tilde{k}_3}{k_t}\right)\xrightarrow{\tilde{k}_3\rightarrow0} -\frac{1}{\mathcal{N}_{\rm nBD}}\frac{\lambda^3\sin{\delta}}{48}\frac{k_1k_2}{k_3^2}.
            \label{eq: flatf}
        \end{equation}
         The normalization factor $\mathcal{N}_{\rm nBD}$, defined in Eq.~\eqref{eq:normfactor}, scales as $\lambda^3$, ensuring an $\mathcal{O}(1)$ shape function in the folded regime. Compared to the equilateral limit of this shape function, $S_{\rm nBD}(k,k,k)\sim3 f(\lambda/3)/{\mathcal{N}_{\rm nBD}}\sim\lambda^2/\mathcal{N}_{\rm nBD}$, the folded limit is point-wise enhanced by $\lambda$.\footnote{The folded-enhanced term is the $\mathcal{O}(|\beta|)$ correction to the usual equilateral- and orthogonal-type enhancement in the bispectrum generated by the EFTofI operator $\dot\zeta^3$~\cite{Chen:2005fe}. Near the folded limit, however, the $\mathcal{O}(|\beta|)$ term is enhanced by $\lambda^3$. Thus, even for $|\beta|\sim \lambda^{-1}$, the folded correction can dominate over the standard equilateral enhancement.} This characteristic folded enhancement in the primordial bispectrum can leave observable traces in the galaxy bispectrum and power spectrum. 
         The enhancement in the folded limit can be understood from the in-in integral Eq.~\eqref{eq:nBDint}. When $\tilde k_jc_s\gtrsim1/|\eta_0|$, the oscillatory phase $\tilde k_jc_s\eta'$ effectively cuts off the integral at $\eta'\sim1/(\tilde k_jc_s)$, leading to
        \begin{equation}
        \int_{\eta_0}d\eta'\eta'^2 e^{i\tilde{k}_jc_s\eta'} \sim\int_{-1/(\tilde{k}_jc_s)} d\eta'\eta'^2\sim\frac{1}{(\tilde k_jc_s)^3}.
        \end{equation}
        As $\tilde k_jc_s$ approaches $0$, the integral has a power-law enhancement.  
        For $\tilde k_jc_s\lesssim1/|\eta_0|$, the integral is cut off at $\eta'\sim\eta_0$ instead:
        \begin{equation}
       \int_{\eta_0}d\eta'\eta'^2 e^{i\tilde{k}_jc_s\eta'} \sim\int_{\eta_0} d\eta'\eta'^2\sim \eta_0^3,
       \end{equation}
       giving the $\eta_0^3\sim\lambda^3$ enhancement in the folded limit as in Eq.~\eqref{eq: flatf}. In the standard Bunch-Davies case, where the phase is $e^{ik_tc_s\eta'}$, we instead find a rapidly oscillating integral that is not enhanced in the folded regime.
       
       The width of the enhanced folded regime will turn out to be important for galaxy clustering constraints. From the condition $\tilde k_jc_s\lesssim1/|\eta_0|$, the cutoff-regulator leads to an enhanced core with a width that scales as $1/\lambda$:
       \begin{equation}
       \tilde k_jc_s\lesssim\frac{1}{|\eta_0|}\Rightarrow
       \tilde k_j\lesssim\frac{1} {\lambda}k_t.
       \label{eq:nBDwidth}
       \end{equation}
        \item \textbf{Squeezed}:
        The behavior of the shape function in the squeezed limit ($k_l/k_s\rightarrow0$, where $k_s$ is a short-wavelength mode and $k_l$ is a long-wavelength mode) is also important for galaxy survey constraints. For example, for local non-Gaussianity, the shape function diverges as $k_s/k_l$ in the squeezed limit, resulting in large-scale enhancement in galaxy power spectrum through scale-dependent bias, which has enabled powerful LSS constraints. 
        
        In the non-Bunch-Davies case analyzed here, due to the presence of another small parameter $1/\lambda$, the squeezed behavior depends on the hierarchy of $k_l/k_s$, $\lambda^{-1}$ and $1$.
        For $k_l/k_s\ll \lambda^{-1}\ll1$, the squeezed behavior satisfies the Maldacena consistency condition~\cite{Maldacena:2002vr, Cheung:2007sv, Flauger:2013hra}:  the shape function decays as $k_l/k_s$, analogous to the equilateral shape function:
        \begin{equation}
        S_{\rm nBD}(|\mathbf{k}_s+\mathbf{k}_l/2|,|\mathbf{k}_s-\mathbf{k}_l/2|,k_l)\xrightarrow{k_l/k_s\ll\lambda^{-1}\ll1}\frac{1}{\mathcal{N}_{\rm nBD}}\frac{1}{24}(3f(\lambda)-\lambda^3\sin{\delta})\left(\frac{k_l}{k_s}\right).
        \end{equation}
        Modes accessible to galaxy surveys, however, typically fall in the moderately-squeezed regime $\lambda^{-1} \ll k_l/k_s\ll 1$, which dominates the late-time structure formation through scale-dependent bias. Averaging over $\mu\equiv\hat{\bk}_l\cdot\hat{\bk}_s$, we find the shape function
        \begin{equation}
            \int^1_{-1}d\mu\ S_{\rm nBD}(|\mathbf{k}_s+\mathbf{k}_l/2|,|\mathbf{k}_s-\mathbf{k}_l/2|,k_l)\xrightarrow{\lambda^{-1}\ll k_l/k_s\ll1}\frac{1}{\mathcal{N}_{\rm nBD}}\frac{\lambda^2}{8}(-\cos(\delta)+\mathcal{O}(k_l/k_s)).
            \label{eq:modsqueezed}
        \end{equation}
        Therefore, the shape function demonstrates a constant scaling behavior in the moderately squeezed regime, sitting between the equilateral and local type~\cite{Flauger:2013hra, Schmidt:2010gw}. The resulting scale-dependent bias is discussed in Sec.~\ref{sec3}.
\end{itemize}

Finally, we note that an upper bound on $\lambda$ is given by back-reaction constraints: the energy density in the excited states, $\rho_{\rm nBD}$ should not exceed the background energy density during inflation, or it will affect the background evolution~\cite{Flauger:2013hra}. For the NPH scenario discussed here, we have
\begin{equation}
\rho_{\text{nBD}}
\sim
\frac{1}{a^4}
\int^{a\Lambda_c/c_s}_{aH/c_s} \frac{d^3k}{(2\pi)^3}
\, |\beta_k|^2 \, c_sk <3M_{\text{pl}}^2H^2\quad\Rightarrow\quad\lambda\lesssim (24\pi^2)^{1/4}c_s^{3/4}|\beta|^{-1/2}\sqrt{\frac{M_{\text{pl}}}{H}}\lesssim c_s^{3/4}\times10^5,
\end{equation}
where in the last step we used $H/M_{\rm pl}\lesssim10^{-5}$ and $|\beta_k|=|\beta|\sim H/M_{\rm pl}$. 

\subsubsection{Imaginary Speed of Sound}
\noindent A second realization of folded enhancement arises in multi-field inflation with strongly non-geodesic motion. In negatively curved field spaces, slow-roll trajectories can be geometrically destabilized and attracted to rapidly turning solutions, leading to a transient tachyonic instability of entropic fluctuations~\cite{Renaux-Petel:2015mga, Cremonini:2010ua, Brown:2017osf, Mizuno:2017idt, Bjorkmo:2019aev, Christodoulidis:2019mkj, Christodoulidis:2019jsx, Aragam:2019omo, Garcia-Saenz:2025jis}. When there is a hierarchy of scales, for example when the absolute value of the mass of the entropic field is well above the Hubble scale, the destabilized heavy entropic field can be integrated out, enabling an effective single-field treatment with an imaginary speed of sound ($c_s^2<0$)~\cite{Garcia-Saenz:2018vqf, Fumagalli:2019noh, Bjorkmo:2019qno, Ferreira:2020qkf, Garcia-Saenz:2025jis, Iarygina:2023msy}.\footnote{It has been recently pointed out that UV matching of such effective treatments needs to be careful to acquire a consistent treatment across all kinematic regimes~\cite{Aoki:2026qea}. However, we find that the distinction is negligible for the forecast setup used in this paper, and leave the study of a full UV model to future work.} Specifically, following~\cite{Garcia-Saenz:2018vqf}, the positive and negative frequency mode functions become exponentially growing and decaying mode functions in the imaginary speed-of-sound case (cf.\,\eqref{eq:MSsol}),
\begin{equation}
    u^+_k=\frac{e^{|c_s|k\eta}}{\sqrt{2|c_s|k}}\left(1-\frac{1}{|c_s|k\eta}\right),\quad u^-_k=\frac{e^{-|c_s|k\eta}}{\sqrt{2|c_s|k}}\left(1+\frac{1}{|c_s|k\eta}\right),
    \label{eq:csmode}
\end{equation}
where $|c_s|=\sqrt{-c_s^2}$. As in the non-Bunch-Davies scenario, the EFT is assumed to be valid only after a mode-dependent time $\eta_0(k)$, which corresponds to the onset of the tachyonic instability for that mode~\cite{Garcia-Saenz:2018vqf}. Here, $\eta_0(k)$ is chosen such that $-k|c_s|\eta_0(k)>\lambda$, with $\lambda\equiv |m_s|/H\gg1$ denoting the ratio of the entropic mass and the Hubble parameter.

In the in-in formalism, the contribution of the operator $\dot\zeta^3$ can be obtained by inserting the mode functions in Eq.~\eqref{eq:csmode}:
\begin{equation}
    2\text{Im}\left[\int_{\eta_0}^\eta d\eta'\langle0|\zeta(\bk_1,\eta')\zeta(\bk_2,\eta')\zeta(\bk_3,\eta')H_{I}(\eta')|0\rangle\right]\propto\sum_{j=1,2,3}\int_{\eta_0}^{\eta}d\eta'\eta'^2 e^{\tilde{k}_j|c_s|\eta'},
    \label{eq:imagcsint}
\end{equation}
which is similar to the conformal time integral of the non-Bunch-Davies case in Eq.~\eqref{eq:nBDint}, where the imaginary exponent becomes real. Again, one can expect an enhancement in the folded regime: the exponent is exponentially suppressed for large $\eta_0$, except for $\tilde k_j\lesssim k_t/\lambda$, when the integral becomes enhanced by $\eta_0^3\sim\lambda^3$. The resulting shape function for this operator is:\footnote{A similar calculation can be carried out for the term $\dot{\zeta}(\partial\zeta)^2$~\cite{Garcia-Saenz:2018vqf}, which features a similar folded enhancement, and a free amplitude.
For the purpose of this paper, we focus on the operator $\dot\zeta^3$ only.}
\begin{equation}
    S_{{\rm imag-}c_s}(k_1,k_2,k_3)=\frac{1}{\mathcal{N}_{{\rm imag-} c_s}}\left\{-\frac{k_1k_2k_3}{(k_1+k_2+k_3)^3}+\sum_{i=1,2,3}\frac{k_1k_2k_3}{\tilde{k}_i^3}\left[1-e^{-\lambda\tilde{k}_i/k_t}\left(1+\lambda\frac{\tilde k_i}{k_t}+\frac{\lambda^2}{2}\frac{\tilde k_i^2}{k_t^2}\right)\right]\right\}.
    \label{eq:icsshape}
\end{equation}
The normalization factor is $\mathcal{N}_{{\rm imag-} c_s}=\lambda^3/192$, again chosen to ensure an $\mathcal{O}(1)$ folded limit. As before, we will require the behavior of the templates in two kinematic limits:
\begin{itemize}
    \item {\textbf{Folded}}: The enhanced piece of the template in the folded regime inherits the scalings discussed above:
    \begin{equation}
    S_{{\rm imag-} c_s}(k_1,k_2,k_3)\xrightarrow{\tilde{k}_3\rightarrow0}\frac{1}{\mathcal{N}_{{\rm imag-} c_s}}\frac{\lambda^3}{48}\frac{k_1k_2}{k_3^2},
    \end{equation}
    which is enhanced by $\lambda^3$ compared to the equilateral regime of $S_{{\rm imag-} c_s}(k,k,k)\sim 1/\mathcal{N}_{{\rm imag-} c_s}$, and has a width of $1/\lambda$ as in Eq.~\eqref{eq:nBDwidth}.
    \item {\textbf{Squeezed}}: When $k_l/k_s\ll \lambda^{-1}\ll1$, the shape function decays as
    \begin{equation}
        S_{{\rm imag-} c_s}(|\mathbf{k}_s+\mathbf{k}_l/2|,|\mathbf{k}_s-\mathbf{k}_l/2|,k_l)\xrightarrow{k_l/k_s\ll\lambda^{-1}\ll1}\frac{1}{\mathcal{N}_{{\rm imag-} c_s}}\frac{\lambda^3-6}{24}\left(\frac{k_l}{k_s}\right),
    \end{equation}
    satisfying the consistency relation. 
    In the moderately squeezed regime, $\lambda^{-1} \ll k_l/k_s\ll 1$, the angularly averaged template demonstrates a constant scaling:
    \begin{equation}
        \int^1_{-1}d\mu\ S_{{\rm imag-} c_s}(|\mathbf{k}_s+\mathbf{k}_l/2|,|\mathbf{k}_s-\mathbf{k}_l/2|,k_l)\xrightarrow{\lambda^{-1}\ll k_l/k_s\ll1}\frac{1}{\mathcal{N}_{{\rm imag-} c_s}}\frac{\lambda^2}{8}(1+\mathcal{O}(k_l/k_s)),
    \end{equation}
    similar to the non-Bunch-Davies case.
\end{itemize}

\subsection{Dissipation-Regulated Folded Enhancement}
\noindent During inflation, the inflaton need not be an isolated degree of freedom, but may interact continuously with other fields in the early universe. A well-known example is warm inflation, in which the inflaton dissipates energy into a radiation bath, whose thermal fluctuations can source curvature perturbations and non-Gaussianity~\cite{Berera:1995ie,Berera:1995wh,Berera:1996nv,Berera:1999ws}. Early studies of warm inflation often start with a Langevin-type equation of motion, with environmental noise sourcing fluctuations and dissipative terms damping their evolution~\cite{LopezNacir:2011kk,LopezNacir:2012rm,Bastero-Gil:2011rva,Bastero-Gil:2014raa}. Microscopic realizations of warm inflation have also been proposed. For example, an inflaton coupled to a non-Abelian gauge sector through axion-like coupling can generate a thermal bath, giving rise to thermal friction and thermal fluctuations during inflation~\cite{Berghaus:2019whh}. A top-down effective-field-theory description of such a scenario has been developed to derive observable predictions~\cite{Mirbabayi:2022cbt}.

Recently, the open effective field theory of inflation (Open EFTofI) has been derived as a systematic bottom-up framework for studying the dynamics of dissipative interactions during inflation (including warm inflation)~\cite{Salcedo:2024smn, Salcedo:2026sdn}. The framework yields predictions for the primordial power spectrum and bispectrum in terms of a set of effective couplings. In the weak-dissipation regime, a characteristic prediction is an enhanced but finite primordial bispectrum near the folded kinematic limit. In this section, we outline the general procedure of calculating the bispectrum in Open EFTofI and analyze the feature of this folded enhancement in de Sitter space. Unlike the previous two examples, the in-in integral is not regulated by a finite initial-time cutoff; instead, dissipation acts as the regulator, suppressing long-time memory and resulting in an exponentially narrow folded core, see Eq.~\eqref{eq:dissiwidth}. 

In the Open EFTofI treatment, the Goldstone mode is doubled into retarded and advanced fields, $\pi_r$ and $\pi_a$. The environmental degrees of freedom are integrated out, leaving an effective open-system description of the inflaton fluctuations. The effective action is then constructed with the allowed terms constrained by the residual symmetries, locality, and a set of non-equilibrium constraints. For example, the quadratic action for canonically-normalized variables reads~\cite{Salcedo:2024smn}
\begin{align}
S_{\mathrm{eff}}^{(2)} &= \int \mathrm{d}^4x \Big\{ a^2 \pi_r' \pi_a' - c_s^2 a^2 \partial_i \pi_r \partial^i \pi_a 
- a^3 \gamma \pi_r' \pi_a + i \Big[ \beta_1 a^4 \pi_a^2 - (\beta_2 - \beta_4) a^2 {\pi_a'}^2 + \beta_2 a^2 (\partial_i \pi_a)^2 \Big] \Big\},
\end{align}
where $\gamma$ represents dissipation effects and $\beta_1$ represents environmental noise. With the action at hand, the curvature power spectrum and bispectra can be calculated through the standard in-in formalism. It is found that in the strong-dissipation regime ($\gamma/H\gg 1$), the bispectra of different operators have enhancement in the equilateral regime, while for weak dissipation ($\gamma/H\ll 1$), the bispectra feature an enhancement in the folded regime. For example, in de Sitter space, the bispectrum generated by the cubic operator $\pi_a^3$ is given by,
\begin{equation}
    B^{\pi_a^3}_\zeta(k_1, k_2, k_3) = -6 \frac{H^3}{f_\pi^6} \frac{\delta_1}{f_\pi^2} \int_{-\infty}^{0} \frac{\mathrm{d}\eta}{H^4\eta^4} G^R(k_1, 0, \eta)G^R(k_2, 0, \eta)G^R(k_3, 0, \eta).
\end{equation}
In this expression, $G^R(k,0,\eta)$ is the retarded Green's function \cite{Salcedo:2024smn}:
\begin{equation}
    G^R(k,0,\eta)=-\frac{H^2}{2k^3}z^3\left(\frac{z}{2}\right)^{-\nu_\gamma}\Gamma(\nu_\gamma)J_{\nu_\gamma}(z)\xrightarrow{z\rightarrow+\infty} \frac{\cos\left(k\eta-\frac{\pi\nu_\gamma}{2}-\frac{\pi}{4}\right)}{(k\eta)^{\frac{\gamma}{2H}-1}},
\end{equation}
where $z=-k\eta$, $\nu_\gamma=\frac{3}{2}+\frac{\gamma}{2H}$. Writing the cosine as an exponential, the correlation function integral asymptotes to
\begin{equation}
    \int_{-\infty}^{\eta_*} d\eta\frac{e^{-i(\pm k_1\pm k_2\pm k_3)\eta}}{(\eta/\eta_*)^{1+\frac{3}{2}\frac{\gamma}{H}}},
    \label{eq:dissiint}
\end{equation}
where the mode mixing in the exponent is sourced by $J_{\nu_\gamma}(z)$, and $\eta_*$ is a reference scale above which the asymptotic form is valid. We thus get a similar mixing of positive- and negative-frequency modes as in Eq.~\eqref{eq:nBDint} and Eq.~\eqref{eq:imagcsint}, but this time the divergence associated with taking the integral limit to $-\infty$ is regulated by the factor $3\gamma/(2H)$ on the power, instead of a cutoff scale. 

\begin{figure}[!tp]
    \includegraphics[width=0.95\linewidth]{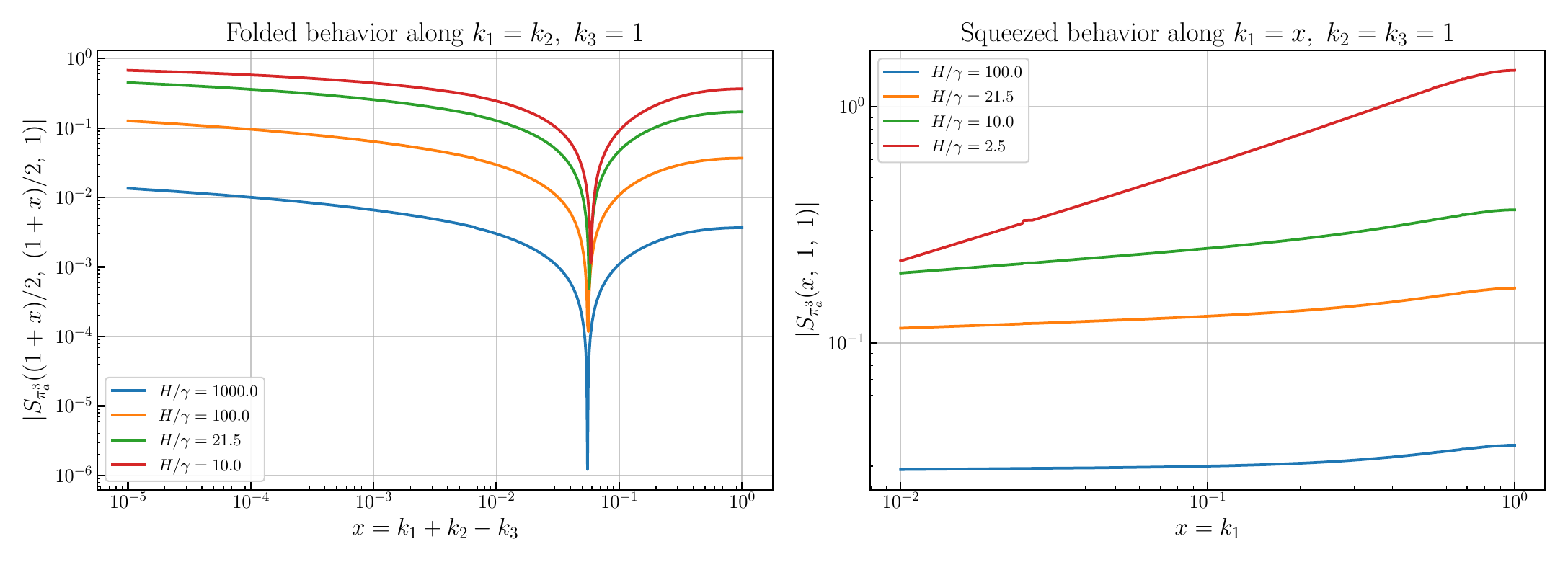}
    
    \caption{Folded and squeezed behavior of the shape function generated by the operator $\pi_a^3$ in dissipative inflation. \textit{Left}: Folded behavior for different $H/\gamma$. 
    With our normalization, all curves approach $S=1$ in the exact folded limit, $x\to0$. Since the saturated folded-enhanced core is exponentially narrow, most of the resolved folded regime is controlled by the logarithmic shoulder. \textit{Right}: Squeezed behavior for different $H/\gamma$.  The squeezed limit of the shape function follows the scaling $(k_l/k_s)^{\gamma/H}$ for $\gamma/H\lesssim 1$.}
    \label{fig:disssquee}
\end{figure}

We define the shape function and the normalization in the same fashion as the previous cases. Since there is no analytical formula for arbitrary operators and values of $\gamma/H$, we focus on the bispectrum induced by the stochastic noise operator $\pi_a^3$ and calculate the shape function numerically (verifying consistency with the results of \citep{Salcedo:2024smn}). As before, we analyze the behavior of the shape function in two kinematic regimes, focusing on the weak-dissipation regime, $\gamma/H\ll1$, where the folded enhancement is most prominent: 
\begin{itemize}
    \item \textbf{Folded}:
        For terms including $e^{-i\tilde{k}_j\eta}$, the integral is effectively cut off at $\eta\sim-1/\tilde k_j$ by the oscillatory exponential. In this case, the integral in Eq.~\eqref{eq:dissiint} can be approximated as:
    \begin{equation}
        \int_{-\infty}^{\eta_*} d\eta\frac{e^{-i\tilde k_j\eta}}{(\eta/\eta_*)^{1+\frac{3}{2}\frac{\gamma}{H}}}\sim\int^{\eta_*}_{-1/\tilde k_j} d\eta\frac{1}{(\eta/\eta_*)^{1+\frac{3}{2}\frac{\gamma}{H}}}=\frac{2H}{3\gamma}\left(1-e^{-3\gamma/(2H)\log( 1/\tilde k_j|\eta_*|)}\right).
        \label{eq:dissienhc}
    \end{equation}
    In the folded regime, the shape is enhanced by $H/\gamma$ compared to the equilateral regime, which sets the normalization factor $\mathcal{N}_{\rm dissi}\propto H/\gamma$. Notably, this enhancement is a generic feature of classically excited fluctuations~\cite{Jiang:2015hfa,Green:2020whw}, here sourced by stochastic environmental fluctuations and regulated by dissipation~\cite{Mirbabayi:2022cbt,Salcedo:2024smn}.
    
    In contrast to the previous cases, the width of the enhanced folded core is exponentially suppressed:
    \begin{equation}
        \log\frac{1}{\tilde k_j}\gtrsim \frac{2H}{3\gamma}\rightarrow \tilde k_j|\eta_*|\lesssim e^{-\frac{2H}{3\gamma}},
        \label{eq:dissiwidth}
    \end{equation}
    with the majority of the folded regime exhibiting only logarithmic enhancement, $\sim \log(1/\tilde k_j|\eta_*|)+\mathcal{O}(\gamma/H)$, from Taylor expanding Eq.~\eqref{eq:dissienhc}. In Fig.~\ref{fig:disssquee}, it can be seen that the shape functions for different $\gamma/H$ experience a vast logarithmic shoulder before reaching the folded core normalized to $\mathcal{O}(1)$. For galaxy surveys, exponentially small $\tilde k_j$ will be hard to reach due to the finite survey volume. For example, setting $\gamma/H\lesssim0.1$ and using values of $\tilde{k}_j$ appropriate for the forecast setup used in Sec.~\ref{sec4}, the $H/\gamma$ enhanced folded core cannot be reached and the enhancement is restricted to $\log(1/\tilde k_j\eta_*)\sim\mathcal O(10)$. \footnote{The exponential narrowing is a consequence of the dissipation regulator, rather than the $1/\eta$ power appearing in Eq.~\eqref{eq:dissiint} (which differs from the $\eta^2$ power in Eq.~\eqref{eq:nBDint}). For example, the operator $\pi_a'^3$ in the Open EFTofI, which carries a scaling $\eta^{2-3\gamma/2H}$, has an asymptotic folded form similar to Eq.~\eqref{eq:dissienhc} :
    \begin{equation}
        \int_{-\infty}^{\eta_*} d\eta\, (\eta/\eta_*)^{2-\frac{3}{2}\frac{\gamma}{H}}e^{-i\tilde k_j\eta}\sim\int^{\eta_*}_{-1/\tilde k_j} d\eta\, (\eta/\eta_*)^{2-\frac{3}{2}\frac{\gamma}{H}}\sim\frac{1-e^{-(-3+3\gamma/(2H))\log(\tilde k_j|\eta_*|)}}{-3+\frac{3}{2}\frac{\gamma}{H}}.
    \end{equation}
    }
    \item \textbf{Squeezed}: 
        We can also look at the squeezed behavior of the shape function generated by the operator $\pi_a^3$ in the weak dissipation regime:
    \begin{equation}
        S_{\rm dissi}(|\mathbf{k}_s+\mathbf{k}_l/2|,|\mathbf{k}_s-\mathbf{k}_l/2|,k_l)\xrightarrow{k_l/k_s\ll1}\frac{1}{\mathcal{N}_{\rm dissi}}\left(\frac{k_l}{k_s}\right)^{\gamma/H},
        \label{eq:dissisquee}
    \end{equation}
    which vanishes as $k_l/k_s\rightarrow 0$ for positive $\gamma/H$. This corresponds to $B_\zeta(k_l,k_s,k_s)/(P_\zeta(k_s)P_\zeta(k_l))\sim(k_l/k_s)^{1+\gamma/H}$ for $k_l\ll k_s$.\footnote{Typically, single-field inflationary models have a squeezed limit scaling as $B_\zeta(k_l,k_s,k_s)/(P_\zeta(k_s)P_\zeta(k_l))\to\mathcal{O}((k_l/k_s)^{2})$~\cite{LopezNacir:2012rm, Maldacena:2002vr, Cheung:2007sv}. This scaling can be different in the case of warm inflation, as noted in~\cite{Mirbabayi:2022cbt}, though the $k_l/k_s\to 0$ limit is still required to vanish. Since (a) the squeezed scaling mainly affects the scale-dependent bias term in the power spectrum, and (b) the constraining power mainly comes from the bispectrum (see the Fisher forecast in Sec.~\ref{sec4}), we leave an analysis on the squeezed limit of the total bispectrum induced by the full set of operators in dissipative inflation to future work.}

\end{itemize}

\subsection{Summary}

\noindent Above, we have considered three examples of models producing folded non-Gaussianity, which fall into two classes according to how the folded-enhanced conformal time integral is regulated and, consequently, how the width of the folded core scales. The common asymptotic in-in integral that results in the folded enhancement is of the schematic form
\begin{equation}
    \int_{-\infty}^{\eta_*}d\eta\,\left(\frac{\eta}{\eta_*}\right)^{p} e^{-i\tilde k_j\eta},
\end{equation}
where $p$ is the power that is specific to the operator chosen. For example, the operator $\dot \zeta^3$ in Eq.~\eqref{eq:nBDint} has $p=2$. For the imaginary-speed-of-sound case, the integral is similar, with the oscillatory exponential replaced by a decaying exponential. As $\tilde k_j\rightarrow0$, this integral is formally divergent for $p\geq-1$. The first kind of regulator is adopted by the non-Bunch-Davies and the imaginary speed-of-sound case, where a cutoff $\eta_0$ to the lower bound of the integral is applied, representing the boundary of the effective description being valid:
\begin{equation}
    \int_{\eta_0}^{\eta_*}d\eta\,\left(\frac{\eta}{\eta_*}\right)^{p} e^{-i\tilde k_j\eta}\sim\left(\frac{\eta_0}{\eta_*}\right)^{p+1}.
\end{equation}
The second kind is adopted by the dissipative case, where the exponentially decaying memory in real time induced by dissipative interactions with the environment translates to a fractional power $p-\alpha$ in conformal time $\eta$:
\begin{equation}
    \int_{-\infty}^{\eta_*}d\eta\,(\eta/\eta_*)^{p-\alpha} e^{-i\tilde k_j\eta}\sim \frac{1-(\tilde k_j|\eta_*|)^{-(p+1-\alpha)}}{-(p+1-\alpha)}.
\end{equation}
For $p=-1$, the originally divergent integral is now regulated since $p+1-\alpha<0$. 

The cutoff-regulated non-Bunch-Davies and imaginary speed-of-sound templates have folded cores whose widths shrink as $1/\lambda$. By contrast, the dissipation-regulated template has a folded core whose width shrinks exponentially, as $e^{-H/\gamma}$. We thus predict that it will be much harder to constrain the exponentially narrow folded-enhanced core of dissipative inflation. The folded and squeezed behaviors of the different scenarios are summarized in Table~\ref{table:folded}:

\begin{table}[h]
\centering
\begin{tabular}{@{}l|cccc@{}}
\toprule
\textbf{Mechanism} 
& \textbf{Folded enhancement\quad} 
& \textbf{Folded core width\quad}  
& \textbf{Squeezed scaling\quad} 
& \textbf{Scale-dep. bias}\\ 
\midrule
non-Bunch-Davies\quad
& $\sim\lambda^{3}$ 
& $\sim \lambda^{-1}$ 
& $\text{const.}\ \to\ k_l/k_s$ 
& $k^{-1}\rightarrow \rm {const.}$\\

imaginary $c_s$
& $\sim\lambda^{3}$ 
& $ \sim \lambda^{-1}$ 
& $\text{const.}\ \to\ k_l/k_s$ 
& $k^{-1}\rightarrow \rm {const.}$\\

dissipative $\pi_a^3$ 
& $\sim H/\gamma$ 
& $\sim e^{-H/\gamma}$ 
& $(k_l/k_s)^{\gamma/H}$ 
& $k^{\gamma/H-1}$\\

\bottomrule
\end{tabular}
\caption{Comparison of folded-enhanced shape functions. The first two columns show the enhancement of the shape function in the folded regime before normalization and the width over which it occurs. The third column shows the squeezed-limit behavior transitioning from moderately squeezed to ultra-squeezed regimes, which is reflected in the low-$k$ scaling of the scale-dependent bias, $\Delta b_1(k)$ (dropping transfer-function corrections $\sim \mathcal{M}(k)/k^2$).}
\label{tab:folded-mechanisms}
\label{table:folded}
\end{table}

\section{Structure Formation in the Presence of Folded PNG}
\label{sec3}
\noindent In this section, we derive the corrections to the redshift-space tree-level galaxy bispectrum and one-loop galaxy power spectrum arising from primordial non-Gaussianities~\cite{Assassi:2015fma, Assassi:2015jqa}. We will mainly follow the treatment of local PNG in~\cite{Cabass:2022ymb}, while keeping the squeezed-limit behavior of the primordial bispectrum general. For the folded PNG models discussed here, different models have different squeezed-limit behavior (summarized in Table~\ref{table:folded}), thus having different scale-dependent bias. We stress that the resulting EFTofLSS implementation is broadly applicable beyond the models considered here, provided that their squeezed limits are no more singular than local PNG and can be described by the same bias expansion.

\subsection{PNG Related Non-Linearity}

\noindent The presence of primordial non-Gaussianity affects structure formation in three ways~\cite{Assassi:2015fma, Assassi:2015jqa,MoradinezhadDizgah:2020whw}. First, we have a non-zero three-point function of the linear matter overdensity:
\begin{equation}
    \langle\delta^{(1)}(\bk_1)\delta^{(1)}(\bk_2)\delta^{(1)}(\bk_3)\rangle=(2\pi)^3\delta_D(\bk_1+\bk_2+\bk_3) B_{111}(k_1,k_2,k_3),
\end{equation}
where $B_{111}$ is sourced by the primordial bispectrum as 
\begin{equation}
    B_{111}(k_1,k_2,k_3)=\mathcal{M}(k_1)\mathcal{M}(k_2)\mathcal{M}(k_3)B_\zeta(k_1,k_2,k_3),
\end{equation}
defining the transfer function $\mathcal{M}(k)=\sqrt{P_{11}(k)/P_\zeta(k)}$.
Secondly, the primordial bispectrum induces additional loop corrections to the power spectrum via the term $P^{\rm NG}_{12}$, which will be discussed shortly.

Thirdly, primordial non-Gaussianity modulates the correlation between long and short modes, resulting in scale-dependent bias. In EFTofLSS, this corresponds to additional terms in the bias expansion of the galaxy overdensity field~\cite{Dalal:2007cu, Matarrese:2008nc, Desjacques:2008vf, Baldauf:2015vio,Desjacques:2016bnm,MoradinezhadDizgah:2012aya}. From the squeezed behavior of the primordial curvature bispectrum, it is straightforward to show that the leading-order additional bias term induced by local non-Gaussianity is $\delta_g \supset b_\zeta f_{\text{NL}} \zeta$, leading to a $1/k^2$ enhancement on large scales for the galaxy power spectrum. For equilateral and orthogonal non-Gaussianity, the additional bias term is $b_\zeta f_{\text{NL}}(k/k_{\text{NL}})^2 \zeta$. In both cases, the squeezed behavior of the primordial bispectra and its impact on local galaxy formation is encapsulated by a power-law scaling: for folded enhanced bispectra, this could break down due to the presence of an additional small parameter (see Sec.~\ref{sec2}), which leads to different scaling behavior in different regimes (see Table~\ref{table:folded}).

\begin{figure}[!tp]
    \includegraphics[width=0.95\linewidth]{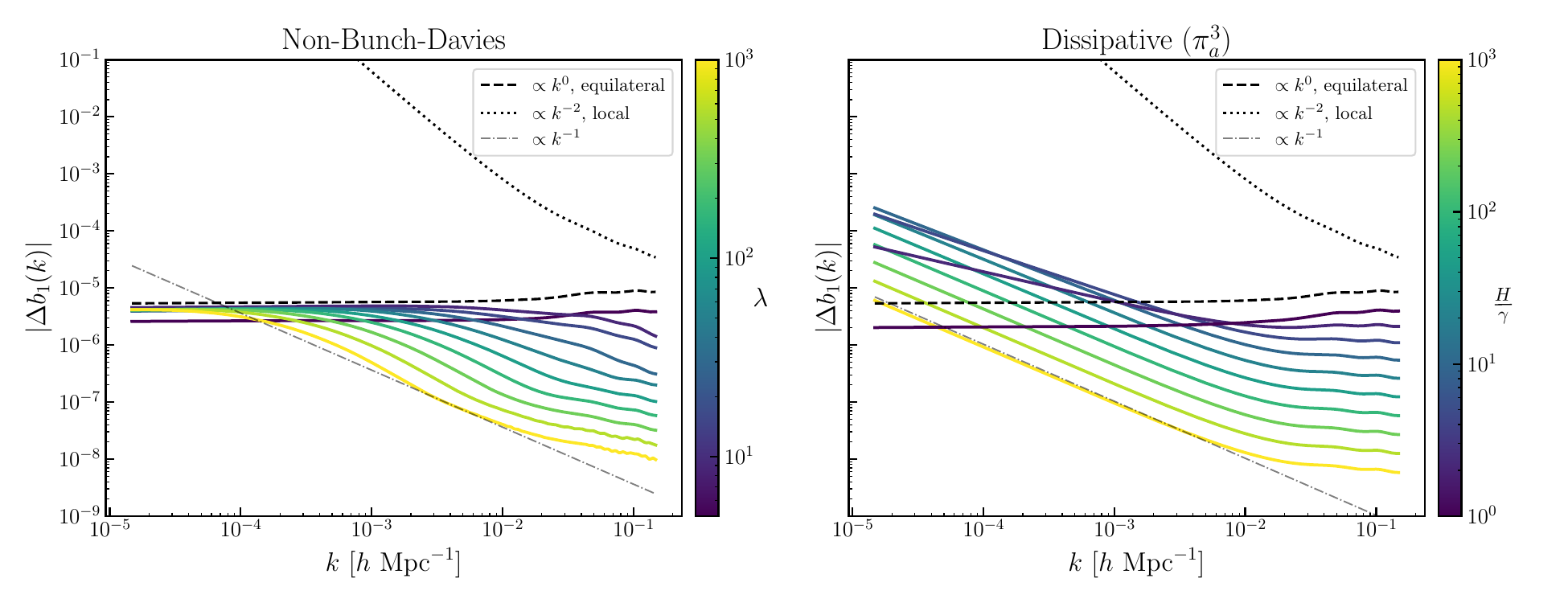}
    \caption{Scale-dependent bias contributions
    $\Delta b_1(k)\propto\mathcal{F}_*^{(3)}(k)/\mathcal{M}(k)$ (plotting the absolute values). \textit{Left}: The non-Bunch-Davies case, for $\lambda\in[1, 10^3]$, representing the cutoff-regulated models. In the moderately squeezed regime, the template sources $|\Delta b_1(k)|\propto k^{-1}$, while at sufficiently small $k$ it transitions to the equilateral-like scaling $k^0$; increasing $\lambda$ moves this transition to lower $k$. \textit{Right}: Dissipative template generated by the operator $\pi_a^3$, for $H/\gamma\in[1, 10^3]$. Its squeezed scaling gives $\Delta b_1(k)\propto k^{\gamma/H-1}$, approaching a $k^{-1}$ enhancement in the weak-dissipation limit. 
    For comparison, we show the biases for equilateral and local non-Gaussianity, which have the conventional low-$k$ scalings of $k^{0}$ and $k^{-2}$.}
    \label{fig:f3}
\end{figure}

A powerful way to estimate the scale-dependent bias generated by a general squeezed
primordial bispectrum was introduced in~\cite{Cabass:2018roz}. This construction smooths the short modes on the characteristic galaxy-formation scale, resulting in a modulation of the local variance that can be converted into a galaxy-abundance response through a peak-background-split argument.  In our analysis we use this construction to capture the scale dependence of the
induced bias, while treating the overall amplitude as a free bias coefficient.  The leading additional bias operator is written as
\begin{equation}
    \delta_g(\bk)
    \supset
    b_\zeta f_{\rm NL}\,
    \mathcal{F}_*^{(3)}(k)\,
    \zeta(\bk),
\end{equation}
where \(b_\zeta\) is an effective bias parameter.  The shape factor
\(\mathcal{F}_*^{(3)}(k)\) is defined as
\begin{equation}
\mathcal{F}_*^{(3)}(k)
=
\frac{1}{4\sigma_*^2 P_\zeta(k)}
\int \frac{{\rm d}^3 k_s}{(2\pi)^3}\,
\mathcal{M}_*(k_1)\,
\mathcal{M}_*(k_2)\,
B_\zeta(k_1,k_2,k)\big|_{f_{\rm NL}=1},
\end{equation}
with $\bk_1=\bk_s-\frac{\bk}{2},\bk_2=-\bk_s-\frac{\bk}{2}$.
The filtered transfer function is $\mathcal{M}_*(k)=\mathcal{M}(k)W_*(k)$, where $W_*(k)=3j_1(kR_*)/(kR_*)$ is the Fourier transform of a spherical top-hat filter with the halo Lagrangian radius $R_*$. In the numerical evaluation, we take $R_*=3.1\,h^{-1}\, \rm Mpc$, corresponding to a halo mass of order $10^{13}\,h^{-1}M_\odot$ \citep{Cabass:2018roz}. $\sigma_*^2$ is the corresponding variance of the density fluctuations smoothed on the scale $R_*$. This expression can be understood as an average of the modes $k_s$ relevant to galaxy formation. The additional operator gives rise to the scale-dependent correction to the linear bias as $b_1\rightarrow b_1+\Delta b_1(k)$, where $\Delta b_1(k)$ is
\begin{equation}
    \Delta b_1(k)
    =
    b_\zeta f_{\rm NL}
    \frac{\mathcal{F}_*^{(3)}(k)}{\mathcal{M}(k)},
\end{equation}
and the scaling of the scale-dependent bias is determined by the squeezed-limit of the shape function:
\begin{equation}
    S(k_s,k_s,k_l)\propto \left(\frac{k_l}{k_s}\right)^N\,\Rightarrow\,\mathcal{F}_*^{(3)}(k)\propto k^{N+1}\,\Rightarrow\,\Delta b_1(k)\propto k^{N-1}\left(\frac{k^2}{\mathcal{M}(k)}\right),
\end{equation}
recalling that $\mathcal{M}(k)\sim k^2$ for $k\lesssim k_{\rm eq}$.

The resulting scale-dependent biases for the different models are shown in Fig.~\ref{fig:f3} and summarized in Table~\ref{tab:folded-mechanisms}. For the cutoff-regulated models,\footnote{Since the imaginary-speed-of-sound model has similar squeezed behavior to the non-Bunch-Davies model, it is omitted from Fig.~\ref{fig:f3}.} the moderately squeezed regime gives $\mathcal{F}_*^{(3)}(k)\propto k$, and hence $\Delta b_1(k)\propto k^{-1}$. At sufficiently small \(k\), the shape function enters the ultra-squeezed regime, resulting in $\mathcal{F}_*^{(3)}(k)\propto k^2$. The corresponding $\Delta b_1$ is then approximately scale independent on sufficiently large scales, as in equilateral PNG. The transition occurs around $k/k_*\sim 1/\lambda$, so increasing $\lambda$ moves the transition to smaller $k$. For the dissipation-regulated model, the squeezed scaling gives $\mathcal{F}_*^{(3)}(k)\propto k^{1+\gamma/H}$ as in Eq.~\eqref{eq:dissisquee}
and therefore $\Delta b_1(k)\propto k^{\gamma/H-1}$. Thus, in the small-$\gamma/H$ limit, the dissipative template approaches a
$1/k$ enhanced scale-dependent bias, matching the squeezed scaling for folded shapes discussed in~\cite{Schmidt:2010gw}.

Following~\cite{Cabass:2022ymb}, the new operator, $f_{\rm NL}\mathcal{F}_*^{(3)}\zeta$, will contribute to a variety of components in the one-loop power spectrum and tree-level bispectrum, both through composite operators such as $f_{\rm NL}\mathcal{F}_*^{(3)}[\zeta\delta]_{\bk}$, redshift-space remapping and advection shifts. These can be packaged in the redshift-space kernels as follows:
\begin{equation}
    Z_1^{\rm NG}(\bk) = f_{\rm NL}\,b_{\zeta}\,\mathcal{F}_*^{(3)}(k)/\mathcal{M}(k)
\end{equation}
for the linear kernel,
\begin{equation}
\begin{split}
    Z_2^{\rm NG}(\bk_1,\bk_2)=&f_{\rm NL}\,b_{\zeta}\,\frac{\bk_1\cdot \bk_2}{2k_1k_2}\left(\frac{k_2}{k_1}\frac{\mathcal{F}_*^{(3)}(k_2)}{\mathcal{M}(k_2)}+\frac{k_1}{k_2}\frac{\mathcal{F}_*^{(3)}(k_1)}{\mathcal{M}(k_1)}\right)\\
    &+f_{\rm NL}b_{\zeta}\,\frac{f\mu k}{2}\left(\frac{\mu_1}{k_1}\frac{\mathcal{F}_*^{(3)}(k_2)}{\mathcal{M}(k_2)}+\frac{\mu_2}{k_2}\frac{\mathcal{F}_*^{(3)}(k_1)}{\mathcal{M}(k_1)}\right)+f_{\rm NL}\,b_{\zeta\delta}\,\frac{1}{2}\left(\frac{\mathcal{F}_*^{(3)}(k_1)}{\mathcal{M}(k_1)}+\frac{\mathcal{F}_*^{(3)}(k_2)}{\mathcal{M}(k_2)}\right)
\end{split}
\end{equation}
for the quadratic kernel, and
\begin{equation}
\begin{split}
    Z^{\rm NG}_3(\bk_1,\bk_2,\bk_3)
= {} & f_{\rm NL}\, b_\zeta
\Bigg[
-\frac{1}{14}\,
\mathcal{G}_2(\bk_1,\bk_2)\,
\frac{(\bk_1+\bk_2)\!\cdot\!\bk_3}{|\bk_1+\bk_2|^2}\,
\frac{\mathcal{F}_*^{(3)}(k_3)}{\mathcal{M}(k_3)}
+ 2~{\rm perms.}
\Bigg]
\nonumber\\
&+ f_{\rm NL}\, b_\zeta
\Bigg[
\frac{1}{6}\,
\frac{\bk_1\!\cdot\!\bk_2\,\bk_2\!\cdot\!\bk_3}{k_1^2 k_2^2}\,
\frac{\mathcal{F}_*^{(3)}(k_3)}{\mathcal{M}(k_3)}
+ 5~{\rm perms.}
\Bigg]
\nonumber\\
&+ f_{\rm NL}\, b_\zeta\,f\mu p_{123}
\Bigg[
\frac{1}{3}\,
\mathcal{G}_2(\bk_1,\bk_2)\,
\frac{\mu_{12}}{|\bk_1+\bk_2|}\,
\frac{\mathcal{F}_*^{(3)}(k_3)}{\mathcal{M}(k_3)}
+ 2~{\rm perms.}
\Bigg]
\end{split}
\end{equation}
for the cubic kernel. Here, $b_\zeta, b_{\zeta\delta}$ are Wilson coefficients, similar to the standard $b_\phi$, $b_{\phi\delta}$ used for local PNG in~\cite{Cabass:2022ymb}.
In these equations $f$ is the logarithmic growth factor, $\hat{\mathbf{z}}$ is the line-of-sight direction unit vector, $\hat{\bk}\equiv \bk/k$, and $\mu_i = \hat{\mathbf{k}}_i\cdot\hat{\mathbf{z}}$. Note that $b_{\zeta\delta^2}$ and $b_{\zeta\mathcal{G}_2}$ do not enter the power spectrum model, as they are removed after renormalization of $b_1$ and $b_\zeta$. 

\subsection{One-Loop Power Spectrum}
\noindent The full one-loop galaxy-galaxy power spectrum in redshift space comprises 
\begin{equation}
    P_{gg}=P_{\rm tree} + P_{22}+ P_{13}+P_{\rm ctr}+P_{\rm stoch}+P^{\rm NG}_{12}+P^{\rm NG}_{\rm tree}+P^{\rm NG}_{22}+P^{\rm NG}_{13},
\end{equation}
where the contributions with superscript $\text{NG}$ indicate order $\mathcal{O}(f_{\rm NL})$ contributions, and the rest are Gaussian contributions. The exact form of the Gaussian pieces can be found in~\cite{Cabass:2022ymb} and references therein.

As discussed in the previous section, there are multiple non-Gaussian contributions. The primordial bispectrum directly gives rise to the loop correction $P^{\text{NG}}_{12}$ through
\begin{equation}
    P^{\rm NG}_{12}(k,\mu)=2Z_1(\bk)\int\frac{d^3q}{(2\pi)^3}Z_2(\bq,\bk-\bq)B_{111}(k,q,|\bk-\bq|),
    \label{eq:PNG12}
\end{equation}
where the kernels can be found in~\cite{Bernardeau:2001qr}. The rest of the terms originate from the scale-dependent bias and are given by:
\begin{equation}
    P_{\rm tree}^{\rm NG}(k,\mu)=2f_{\rm NL}\,(b_1+f\mu^2)\, b_{\zeta}\,\mathcal{F}_*^{(3)}(k)\frac{P_{11}(k)}{\mathcal{M}(k)},
    \label{eq:PNGtree}
\end{equation}
\begin{equation}
    P^{\rm NG}_{22}(k,\mu) = 4\int\frac{d^3q}{(2\pi)^3}\,Z^{\rm NG}_2(\bq,\bk-\bq)Z_2(\bq,\bk-\bq)\,P_{11}(q)P_{11}(|\bk-\bq|),
    \label{eq:PNG22}
\end{equation}
\begin{equation}
\begin{split}
    P^{\rm NG}_{13}(k,\mu)= 6Z^{\rm NG}_1(\bk)P_{11}(k)\int\frac{d^3q}{(2\pi)^3}\,Z_3(\bk,\bq,-\bq)\,P_{11}(q)+6Z_1(k)P_{11}(k)\int\frac{d^3q}{(2\pi)^3}\,Z^{\rm NG}_3(\bk,\bq,-\bq)\,P_{11}(q),
    \label{eq:PNG13}
\end{split}
\end{equation}
where $P_{11}$ is the linear matter power spectrum and the kernels with superscript NG are listed in the previous subsection. 

At one-loop order, the integral proportional to $b_2$ in the PNG contribution $P^{\rm NG}_{12}$ induces a counterterm. However, this counterterm is degenerate with $P_{\rm tree}^{\rm NG}$, which can be seen by taking the large-$q$ limit of the integral \citep[e.g.,][]{Assassi:2015fma}. Therefore, instead of introducing a new counterterm, we absorb its contribution into the renormalized coefficient $b_\zeta$. For $P_{22}^{\rm NG}$ and $P_{13}^{\rm NG}$, the induced counterterms are, in principle:
\begin{equation}
    P_{\rm ctr}^{\rm NG}(k)=-2f_{\rm NL}\sum_{d=0}^2\tilde{c}^{\rm NG}_{2d}\,f^d\mu^{2d}\,\frac{\mathcal{F}_*^{(3)}(k)}{\mathcal{M}(k)}k^2P_{\rm 11}(k).
    \label{eq:PNGctr}
\end{equation}
The loop order of the counterterms depends on the scaling of $\mathcal{F}_*^{(3)}(k)$, and is generally at least one-loop due to the factor $k^2P(k)$. We do not include these PNG counterterms in the forecast. As justified in Sec.~\ref{sec:scaling}, for the folded shapes considered here these terms enter at higher order than the Gaussian loop counterterms. 

So far we have focused on the deterministic part of the galaxy power spectra. There is an extra stochasticity term that needs to be included in the presence of primordial non-Gaussianity: 
\begin{equation}
    \delta_g(\bk) \supset f_{\rm NL}\,b_{\zeta\epsilon}\frac{\mathcal{F}_*^{(3)}(k)}{\mathcal{M}(k)}\,[\zeta\epsilon]_\bk.
    \label{eq:stochterm}
\end{equation}
For power spectrum, however, the non-zero correlations of this additional term with the other operators are either degenerate with counterterms or enter at higher order. Therefore, no new stochastic contribution is needed, as in \citep{Cabass:2022ymb}.

\subsection{Tree-Level Bispectrum}
\noindent The full tree-level bispectrum consists of
\begin{equation}
    B_{ggg}=B_{\rm 211}+B_{\rm stoch}+ B^{\rm NG}_{111}+B^{\rm NG}_{\rm 211},
\end{equation}
where the form of the Gaussian pieces can be found in~\cite{Cabass:2022ymb}.

Similar to the power spectrum, the non-Gaussian contributions come from the primordial bispectrum:
\begin{equation}
B^{\rm NG}_{111}=Z_1(\bk_1)Z_1(\bk_2)Z_1(\bk_3)B_{111}(k_1,k_2,k_3). 
\end{equation}
as well as from the scale-dependent bias kernels,
\begin{equation}
\begin{split}
   B_{211}^{\rm NG}
   =2[Z_1^{\rm NG}(\bk_1)Z_1(\bk_2)Z_2(\bk_1,\bk_2)+Z_1(\bk_1)Z_1^{\rm NG}(\bk_2)Z_2(\bk_1,\bk_2)+Z_1(\bk_1)Z_1(\bk_2)Z_2^{\rm NG}(\bk_1,\bk_2)] P_{L}(\bk_1)P_{L}(\bk_2)+2\,\rm perms. 
\end{split}
\end{equation}

The stochastic term in Eq.~\eqref{eq:stochterm} induces new contributions to the redshift-space bispectrum as 
\begin{equation}
    B^{\rm NG}_{\rm stoch}=f_{\rm NL}(b_\zeta+b_{\zeta\epsilon})\,\frac{\mathcal{F}_*^{(3)}(k_1)}{\mathcal{M}(k_1)}\left[\frac{b_1B_{\rm shot}+(1+P_{\rm shot})f\mu_1^2}{\bar{n}}P_{\rm 11}(k_1)\right] + 2\,\rm perms.
\end{equation}
In~\cite{Cabass:2022ymb}, this contribution was neglected because it is suppressed in the squeezed limit compared to $B_{211}^{\rm NG}$. Since the most constraining power on the folded templates discussed here comes from folded rather than squeezed regimes (see Sec.~\ref{sec4}), we include $B_{\rm stoch}^{\rm NG}$ in the forecast. This term is not itself folded enhanced, but is an additional smooth stochastic PNG contribution.

\subsection{Scaling Behavior}
\label{sec:scaling}

\noindent The relative size of various perturbative corrections can be estimated using the scaling universe approach~\cite{Assassi:2015fma, Cabass:2022ymb,Cabass:2022wjy}. This is based on the observation that the linear matter power spectrum can be well approximated by a power-law $P_{11} \propto (k/k_{\text{NL}})^n k_{\text{NL}}^{-3}$ with an effective spectral index $n \approx -1.7$ for quasi-linear wavenumbers $k \simeq 0.15 \, h\text{Mpc}^{-1}$. Introducing the nonlinear scale $k_{\text{NL}}$, the estimates for the dimensionless galaxy power spectrum $\Delta^2(k) \equiv k^3 P(k) / (2\pi^2)$ for the Gaussian contributions give \citep[e.g.,][]{Cabass:2022ymb}:
\begin{equation}
 \Delta^2_{\rm G}(k) = \underbrace{\left(\frac{k}{k_{\text{NL}}}\right)^{1.3}}_{\text{tree}} + \underbrace{\left(\frac{k}{k_{\text{NL}}}\right)^{2.6}}_{\text{1-loop}} + \underbrace{\left(\frac{k}{k_{\text{NL}}}\right)^{3.3}}_{\text{ctr}} + \underbrace{\left(\frac{k}{k_{\text{NL}}}\right)^{3}}_{\text{stoch}}. 
 \end{equation}
The relative importance of the contributions from primordial non-Gaussianity, $\Delta^2_{\rm NG}$, depends on the scaling of $\mathcal{F}_*^{(3)}(k)$ for the range of modes investigated. For $\mathcal{F}_*^{(3)}(k)\propto k^{N+1}$, 
\begin{equation}
\Delta^2_{\text{NG}}(k) = f_{\text{NL}}  \Delta_\zeta \underbrace{\left(\frac{k}{k_{\text{NL}}}\right)^{(N+1)+0.65}}_{\text{PNG tree}} 
+f_{\text{NL}}  \Delta_\zeta \underbrace{\left(\frac{k}{k_{\text{NL}}}\right)^{1.95}}_{P_{12}} 
+ f_{\text{NL}} \Delta_\zeta \underbrace{\left(\frac{k}{k_{\text{NL}}}\right)^{(N+1)+1.95}}_{\text{PNG scale-dep. loop}}
+ f_{\text{NL}} \Delta_\zeta \underbrace{\left(\frac{k}{k_{\text{NL}}}\right)^{(N+1)+2.65}}_{\text{PNG loop ctr.}},
\end{equation}
where ``PNG scale-dep. loop'' refers to the loop contributions induced by the scale-dependent bias (Eq.~\eqref{eq:PNG22},~\eqref{eq:PNG13}) and ``PNG loop ctr.'' refers to the counterterms induced by those terms (Eq.~\eqref{eq:PNGctr}). For local PNG, $N=-1$, thus all terms are comparable with the Gaussian contributions and need to be included in the analysis~\cite{Cabass:2022ymb}. For equilateral and orthogonal PNG, $N=1$, and we only need to include the tree-level PNG term (Eq.~\eqref{eq:PNGtree}) and $P_{12}$ (Eq.~\eqref{eq:PNG12})~\cite{Cabass:2022wjy}.

In the case of folded non-Gaussianity, the scaling of the terms is modified. As an example, for the non-Bunch-Davies and imaginary speed-of-sound case, we have two different squeezed regimes as summarized in Table~\ref{tab:folded-mechanisms}. When $k/k_*<\lambda^{-1}$, we have $N=1$ and the scale-dependent bias is approximately scale-independent, similar to the case of equilateral PNG. The PNG contributions therefore scale as~\cite{Cabass:2022wjy}:
\begin{equation}
\Delta^2_{\text{NG}}(k) = f_{\text{NL}} \Delta_\zeta \underbrace{\left(\frac{k}{k_{\text{NL}}}\right)^{2.65}}_{\text{PNG tree}} 
+ f_{\text{NL}} \Delta_\zeta \underbrace{\left(\frac{k}{k_{\text{NL}}}\right)^{1.95}}_{P_{12}}
+ f_{\text{NL}} \Delta_\zeta \underbrace{\left(\frac{k}{k_{\text{NL}}}\right)^{3.95}}_{\text{PNG scale-dep. loop}}
+ f_{\text{NL}} \Delta_\zeta \underbrace{\left(\frac{k}{k_{\text{NL}}}\right)^{4.65}}_{\text{PNG loop ctr.}}, 
\end{equation}
where the scale-dependent-bias-induced loop corrections are sub-dominant. However, for large $\lambda$, the observationally accessible modes usually satisfy $k/k_*>\lambda^{-1}$. In this regime, we have $N=0$, thus the scale-dependent bias scales as $\Delta b_1(k)\propto k^{-1}$, and the scaling of $\Delta_{\rm NG}^2(k)$ becomes
\begin{equation}
\Delta^2_{\text{NG}}(k) = f_{\text{NL}} \Delta_\zeta \underbrace{\left(\frac{k}{k_{\text{NL}}}\right)^{1.65}}_{\text{PNG tree}} 
+ f_{\text{NL}} \Delta_\zeta \underbrace{\left(\frac{k}{k_{\text{NL}}}\right)^{1.95}}_{P_{12}}
+ f_{\text{NL}} \Delta_\zeta \underbrace{\left(\frac{k}{k_{\text{NL}}}\right)^{2.95}}_{\text{PNG scale-dep. loop}}
+ f_{\text{NL}} \Delta_\zeta \underbrace{\left(\frac{k}{k_{\text{NL}}}\right)^{3.65}}_{\text{PNG loop ctr.}}, 
\end{equation}
where the scale-dependent-bias-induced loop contributions are less enhanced than the local case, but still comparable to the Gaussian loop contributions. For the weak-dissipation case we have $N=\gamma/H\rightarrow0$, so that the PNG contributions have a similar scaling. As such, we conclude that it is important to include the PNG loops in our analysis, but not the counterterms induced by the PNG loops (which appear at higher order). 

\subsection{Practical Computation}
\noindent The quantities of observational interest are the binned power spectrum and bispectrum multipoles. For the redshift-space power spectra $P(k,\mu)$ and bispectra $B(k_1,k_2,k_3,\mu_1,\mu_2,\mu_3)$ computed using the model in the previous sections, we define the multipoles as usual \citep[e.g.,][]{Scoccimarro:2015bla,Ivanov:2023qzb},
\begin{equation}
    P_\ell(k)=(2\ell+1)\int_{-1}^1\frac{d\mu}{2} \,\mathcal{L}_\ell(\mu)P(k,\mu),
\end{equation}
\begin{equation}
B_\ell(k_1,k_2,k_3)
=
(2\ell+1)\int_{-1}^{1} \frac{d\mu}{2}
\int_{0}^{2\pi} \frac{d\phi}{2\pi}
\,\mathcal{L}_\ell(\mu_1)
B\!\left(
k_1, k_2, k_3,
\mu_1[\mu],
\mu_2[\mu,\phi],
\mu_3[\mu,\phi]
\right),
\end{equation}
where $\mathcal{L}_\ell$ is the Legendre polynomial of order $\ell$, $\mu_1 = \mu, \,\mu_2 = \mu \cos\alpha - \sqrt{1-\mu^2}\,\sin\alpha\,\cos\phi,$
and $\cos\alpha \equiv \hat{\mathbf{k}}_1 \cdot \hat{\mathbf{k}}_2$. In this paper, we will focus on the power spectrum multipoles $\ell=0,2,4$ and the bispectrum monopole, $\ell=0$. For the Gaussian pieces, we use the results computed using \textsc{class-pt}~\cite{Chudaykin:2020aoj}, while for the non-Gaussian pieces, we numerically compute the loop integrals entering the power spectrum and bispectrum multipoles for the folded-PNG models discussed in the previous sections. As a validation, we have checked our implementation of $P_{12}^{\rm NG}$ against \textsc{class-pt} for the equilateral and orthogonal templates \citep{Cabass:2022wjy}, which is based on the \textsc{fft-log} method \citep{2018JCAP...04..030S}.
While much of the calculation must be done numerically (due to the non-trivial shape functions used in this work), part of the redshift-space integration can be performed analytically; this is detailed in Appendix~\ref{app:MD}.

We have implemented infrared (IR) resummation for the PNG terms entering the power spectrum and bispectrum models, following the formalism of time-sliced perturbation theory~\cite{Blas:2016sfa, Oddo:2019run,Vasudevan:2019ewf,Ivanov:2018gjr}.\footnote{Following \citep{Philcox:2022frc,Bakx:2025pop}, we use the angle-averaged damping exponents to compute the PNG loop integrals, considerably simplifying the angular integration. Though the approach is approximate, the error is suppressed by both loops and $f_{\rm NL}$.} Note that for the non-Bunch-Davies case, we do not implement IR resummation for the fast oscillatory features. It would be interesting to implement a rigorous IR resummation for the fast oscillations as in~\cite{Vasudevan:2019ewf}, which is expected to lead to suppressed oscillations on quasi-linear scales. However, as will be shown in Sec.~\ref{sec4}, the constraining power for the non-Bunch-Davies case does not significantly rely on the fast oscillations, thus we leave the rigorous treatment for future study.  

The final step is to average the theoretical spectra over the same Fourier-space bins used in the observation. We use the integration scheme to account for the binning effect \cite{Ivanov:2021kcd}. For a set of linearly-spaced bins with $k_a\in[\bar k_a-\Delta k/2, \bar k_a+\Delta k/2]$, we define the binned power spectrum multipoles as 
\begin{equation}
    P^{\rm bin}_{\ell}(\bar{k}_a)
    =
    \frac{(2\ell+1)}{\mathcal{N}(\bar k_a)}
    \int_{\bar{k}_a-\Delta k/2}^{\bar{k}_a+\Delta k/2} k^2dk\,  \int_0^{2\pi}\frac{d\phi}{2\pi}\int_{-1}^{1}\frac{d\mu}{2}\mathcal{L}_{\ell}(\mu)P(k,\mu),
\end{equation}
where the normalization factor is
\begin{equation}
    \mathcal{N}(\bar k_a)=\int_{\bar k_a-\Delta k/2}^{\bar k_a+\Delta k/2}k^2dk\int_{-1}^{1} \frac{d\mu}{2}
\int_{0}^{2\pi} \frac{d\phi}{2\pi}\,.
\end{equation}
For the bispectrum, we consider only ordered triangle bins whose centers satisfy $\bar k_1\leq \bar k_2\leq \bar k_3\text{ and }\bar k_3\leq \bar k_1+\bar k_2$. The binned theory prediction is computed by a weighted sum of the bispectrum monopole $B_0$ over a set of sampled continuum triangles inside each bin:
\begin{equation}
    B_0^{\rm bin}(\bar k_1,\bar k_2,\bar k_3)
    =
    \frac{
    \sum_{i\in{\rm bin}}
    k_{1,i}k_{2,i}k_{3,i}(\Delta k)^3\,
    \mathcal{I}(k_{1,i},k_{2,i},k_{3,i})\,
    B_0(k_{1,i},k_{2,i},k_{3,i})
    }{
    \sum_{i\in{\rm bin}}
    k_{1,i}k_{2,i}k_{3,i}\Delta k^3\,
    \mathcal{I}(k_{1,i},k_{2,i},k_{3,i})
    }.
\end{equation}
Here $\mathcal{I}=1$ if the sampled side lengths form a triangle and
$\mathcal{I}=0$ otherwise.  The factor \(k_1k_2k_3(\Delta k)^3\) is the continuum
phase-space weight for triangle configurations.\footnote{We use the numerical weighted sum to approximate the continuum bin integral rather than to perform a discrete Fourier-mode
average appropriate for a finite survey volume. We do not include the `discreteness' correction discussed in~\cite{Ivanov:2021kcd} (arising from the latter sum), since \cite{Bakx:2025pop} found this to have a negligible impact on the bispectrum monopole.}

\section{Results and Discussion}
\label{sec4}
\noindent In this section, we estimate the constraining power on folded primordial non-Gaussianity using a Fisher forecast for the redshift-space power spectrum and bispectrum. We use the power spectrum multipoles $P_\ell(k)$ with $\ell=0,2,4$, and the bispectrum monopole summarized in Sec.~\ref{sec3}, which depend on the nuisance parameters
\begin{equation}
    \{b_1,b_2,b_{\mathcal G_2},b_{\Gamma_3},
c_0,c_2,c_4,\tilde c,c_1,
P_{\rm shot},B_{\rm shot},a_0,a_2,b_\zeta,b_{\zeta\delta},b_{\zeta\epsilon}\}.
\end{equation}
These describe the galaxy bias up to third order, redshift-space counterterms, and stochastic contributions. The last three parameters are associated with the PNG-induced scale-dependent bias and stochastic terms. We adopt $f_{\rm NL}^{\rm fid}=0$ and $b_1^{\rm fid}=2.5$. We set $b_2^{\rm fid}$ and $b_{\mathcal{G}_2}^{\rm fid}$ using the simulation-calibrated fitting relations of~\citep{Ivanov:2024xgb}, $b_{\Gamma_3}^{\rm fid}$ using the coevolution relation, $b_{\zeta}^{\rm fid}, b_{\zeta\delta}^{\rm fid}$ using the ``universality'' relations~\citep{Desjacques:2016bnm}, and $b_{\zeta\epsilon}^{\rm fid}=0$. Unless otherwise stated, we marginalize over these nuisance parameters with broad Gaussian priors following~\cite{Chudaykin:2025aux}. For both the power spectrum and bispectrum, we include modes from $k_{\rm min}=0.005\,\hmpc$ to $k_{\rm max}=0.155\,\hmpc$, with bin width $\Delta k=0.005\,\hmpc$, unless otherwise stated. The lower cutoff is chosen to be representative of DESI-like survey volumes~\cite{Chudaykin:2025vdh}, while the upper cutoff is chosen to be in the range where the perturbative modeling is expected to be under reasonable control.\footnote{The cutoff \(k_{\rm max}=0.155\,\hmpc\) is somewhat optimistic for a tree-level bispectrum model, but is close to the scale cuts used in analyses including the one-loop galaxy bispectrum~\cite{Bakx:2025pop}. Since the folded PNG templates considered here are only mildly affected by marginalization over nuisance parameters, we do not expect the qualitative conclusions to change substantially if one instead used the one-loop Gaussian bispectrum and tree-level primordial contribution.} 
The forecast is performed at redshift $z=0$, assuming a survey volume $V=2.5\ h^{-3}\,\rm Gpc^3$ and a galaxy number density $\bar n=4\times10^{-4}\,h^3\,\rm Mpc^{-3}$, similar to the DESI LRG sample \citep{DESI:2024aax}. We assume Gaussian covariances for both statistics~\cite{Chudaykin:2019ock}, and neglect the cross-covariance between the two. All constraints below refer to the normalization convention defined in Sec.~\ref{sec2}, in which the folded core is normalized to $\mathcal{O}(1)$.

\subsection{Cutoff-Regulated Case: Non-Bunch-Davies and Imaginary Speed of Sound}
\noindent The non-Bunch-Davies and imaginary speed-of-sound templates have folded cores whose widths scale as $1/\lambda$, as discussed in Sec.~\ref{sec2}. Below, we study how the folded enhancement and its morphology impact the constraining power. 

\begin{figure}[!tp]
    \centering
    \includegraphics[width=0.95\linewidth]{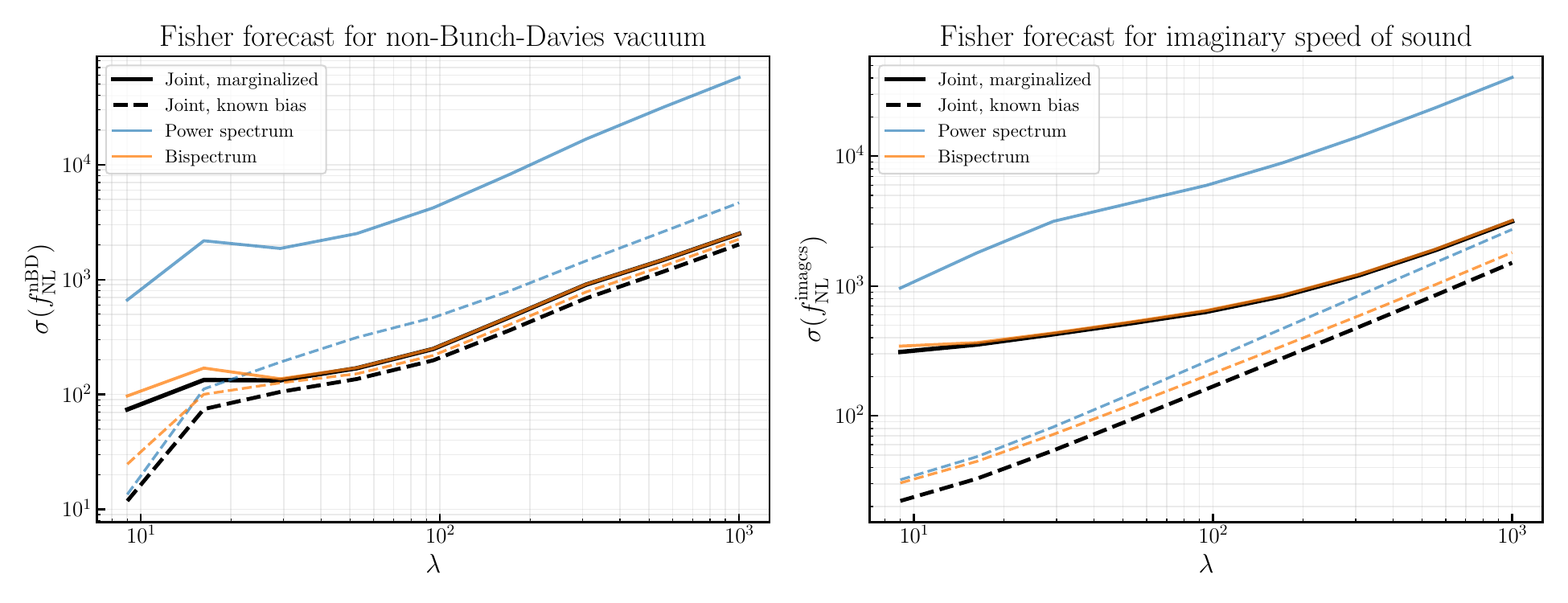}
    \caption{Fisher forecasts on $f_{\rm NL}$ sourced by non-Bunch-Davies vacuum states (left) and imaginary speed-of-sound models (right). In both cases, the constraining power on $f_{\rm NL}$ mostly comes from the bispectrum, due to the characteristic folded enhancement. This signature is very distinct from galaxy formation physics, as can be seen from the relatively small loss of information from marginalizing over nuisance parameters (solid versus dashed lines). The constraints weaken as $\lambda$ increases since the width of the folded enhancement (normalized to $\mathcal{O}(1)$) shrinks with increasing $\lambda$ and makes the signal harder to detect, as explained in the main text and shown in Fig.~\ref{fig:bin}.}
    \label{fig:joint}
\end{figure}

In Fig.~\ref{fig:joint}, the Fisher forecast constraints on folded non-Gaussianity from the galaxy power spectrum, bispectrum, and their combination are displayed. We show the uncertainties on $f_{\rm NL}$ for both the non-Bunch-Davies and imaginary speed-of-sound scenario. In both cases, the bispectrum provides the dominant source of constraining power over most of the parameter range. This is expected, as the galaxy bispectrum contains geometric information on the triangular configurations, receiving tree-level contributions from the folded-enhanced primordial bispectrum. The galaxy power spectrum, however, depends only on a single $k$-mode and therefore has no direct handle on the enhancement in the folded regime; it is sensitive to folded PNG only indirectly, through scale-dependent bias and loop corrections. To assess the importance of the scale-dependent-bias terms, we repeat the forecasts with $b^{\rm fid}_\zeta=b^{\rm fid}_{\zeta\delta}=0$, thereby removing these terms from the response of the spectra to $f_{\rm NL}$. Compared with our fiducial choices as in~\citep{Desjacques:2016bnm}, this changes the marginalized power spectrum constraints by a factor of $2$, while the bispectrum and joint constraints change by less than $1\%$ at large $\lambda$. This indicates that scale-dependent-bias terms carry much of the power spectrum information, whereas the bispectrum information, which dominates overall, is sourced by the direct primordial contribution.

Fig.~\ref{fig:joint} also illustrates the impact of marginalizing over the nuisance parameters. For small $\lambda$, marginalization can noticeably weaken the constraint, as is the case for equilateral and orthogonal shapes~\cite{Cabass:2022epm,Baumann:2021ykm}. This is because the folded templates still have significant overlap with equilateral or orthogonal shapes, and the constraints on $f_{\rm NL}$ are limited by our lack of knowledge of nuisance parameters. However, for larger $\lambda$, the feature becomes increasingly localized along the folded boundary, as shown in Fig.~\ref{fig:EIS}, making it geometrically distinct from the bias contributions. Therefore, marginalization only causes a relatively small loss of information, and the large-$\lambda$ constraints are mainly limited by the information extractable from the folded triangles, rather than by degeneracies with galaxy bias. Importantly, this feature is rather generic for the templates with a folded ridge (as can be seen from the two cases in Fig.~\ref{fig:joint}), with the information loss further reduced by the oscillatory feature in the non-Bunch-Davies model. 

\begin{figure}[!tp]
    \centering
    \includegraphics[width=0.95\linewidth]{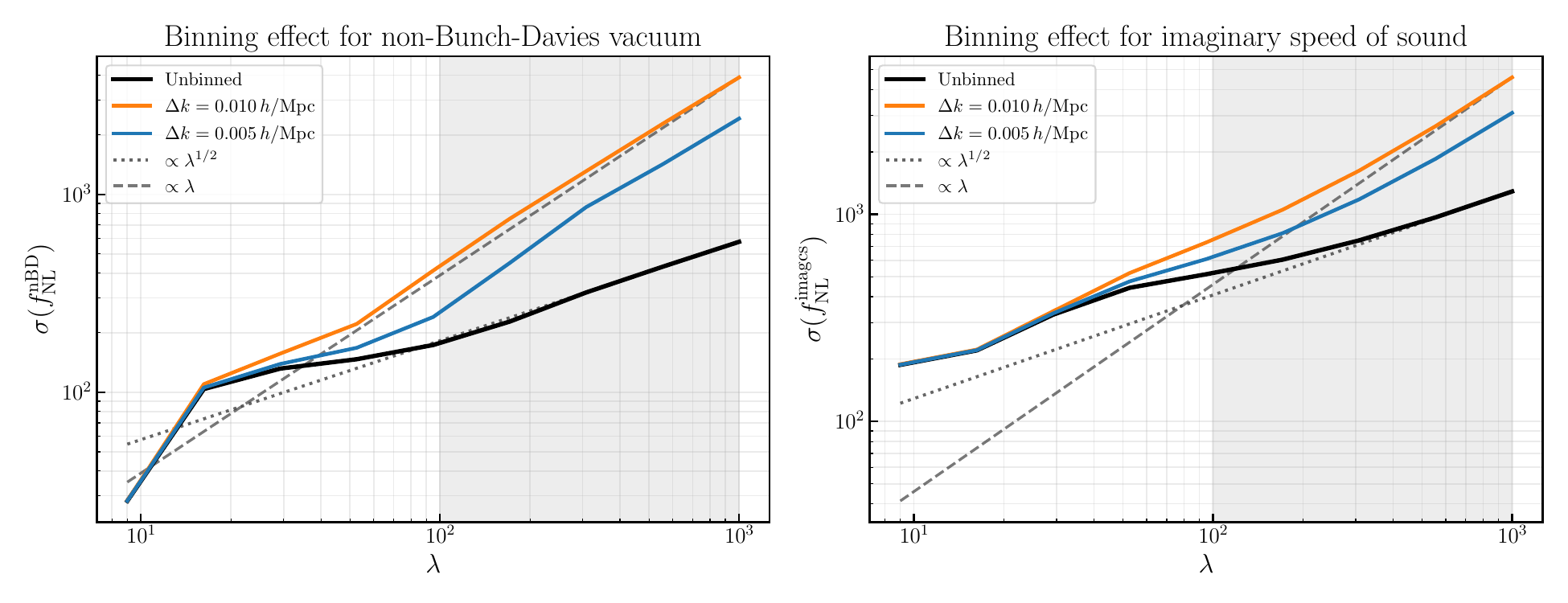}
    \caption{Effect of binning on the PNG Fisher forecasts for the non-Bunch-Davies model (left) and imaginary speed-of-sound scenario (right). We have projected out any equilateral and orthogonal piece from the template for demonstration. The gray region indicates regimes with a significant loss of constraining power due to binning. The unbinned case has a scaling of $\sigma(f_{\rm NL})\propto\sqrt{\lambda}$, which is due to the folded core that has a power-law width $1/\lambda$. The binned forecast agrees with the unbinned case until the folded-core width becomes comparable or smaller than the bin size. For $\lambda\gtrsim10^2$, the bin averaging dilutes the folded signal and the scaling of the binned forecast weakens to $\sigma(f_{\rm NL})\propto \lambda$, implying that highly enhanced configurations will be difficult to detect in practice.} 
    \label{fig:bin}
\end{figure}

As discussed in Sec.~\ref{sec2}, the width of the folded core shrinks as $\lambda^{-1}$. Since any realistic galaxy survey uses finite-width $k$-bins (set by the fundamental mode), it is important to ask when the narrowing folded core becomes unresolved due to binning. Schematically, without binning, the Fisher information $F\sim\sum_k (\partial B/\partial f_{\rm NL})C^{-1}(\partial B/\partial f_{\rm NL})$ is dominated by the localized enhancement in the folded regime, implying that $\sigma(f_{\rm NL})$ scales as:
\begin{equation}
    F^{\rm unbin}\ \sim
    \underbrace{\mathcal O(1)}_{(\text{folded amp.})^2}\times \underbrace{1/\lambda}_{\text{width}}
    \;\;\Rightarrow\;\;
    \sigma(f_{\rm NL}) \propto \sqrt{\lambda}
\end{equation} 
over most of the parameter range. When binning is introduced, the bin width is still smaller than the width of the folded-enhanced core for small $\lambda$, thus the enhancement is fully resolved and the binned constraint will follow $\sigma(f_{\rm NL})\propto \sqrt{\lambda}$. For large $\lambda$, however, the folded core shrinks to be narrower than the bin width, and binning will then dilute the enhanced signal. Since each bin in the folded regime contains $\sim1/\lambda$ number of folded triangles, the averaging gives $\sum_{k\in\rm bin} \partial B / \partial f_{\mathrm{NL}}\sim\mathcal O(1)\times1/\lambda$ (for each individual bin), and the resulting scaling of $\sigma(f_{\rm NL})$ is weakened by $\sqrt{\lambda}$: 
\begin{equation}
    F^{\rm bin}\sim\ \sum_\text{bin}\left[\sum_{k\in\text{bin}} (\partial B/\partial f_{\rm NL})\right]C^{-1}_{\rm bin}\left[\sum_{k\in\text{bin}} (\partial B/\partial f_{\rm NL})\right]\ \sim\frac{1}{\lambda^2}\quad\Rightarrow\quad\sigma(f_{\mathrm{NL}})\sim\lambda\,.
\end{equation}

In Fig.~\ref{fig:bin}, we compare the bispectrum forecast for the unbinned and binned cases for both the non-Bunch-Davies and imaginary sound-speed models. Note that in the figure, we project out the equilateral and orthogonal components from the shape functions to isolate the information associated with the localized folded core. It can be seen that the transition between the two scaling regimes happens at $\lambda\sim10^2$: for $\lambda\lesssim10^2$, the binned curves overlap with the unbinned ones and approximately scale as $\sqrt{\lambda}$. For $\lambda\gtrsim10^2$, the binned curve deviates from the unbinned curve and scales as $\lambda$, implying that binning causes a significant loss of information. 

\subsection{Dissipation-Regulated Case: Dissipative Inflation}
\noindent Dissipative inflation belongs to the second class of models discussed in Sec.~\ref{sec2}. In this case, the folded core is exponentially narrow in the weak-dissipation regime and is therefore observationally hard to access. We use the operator $\pi_a^3$ in the Open EFTofI as a representative example of this behavior. 

\begin{figure}[!tp]
    \centering
    \includegraphics[width=0.95\linewidth]{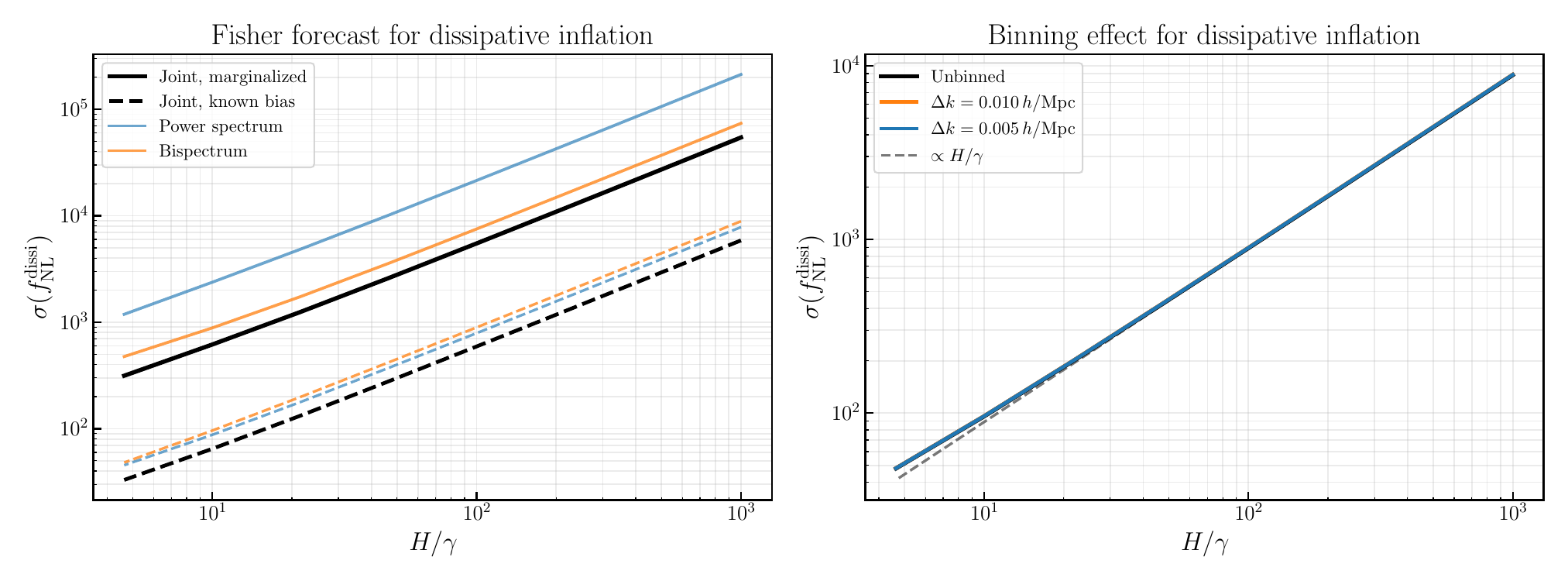}
    \caption{\textit{Left}: Fisher forecast result for the operator $\pi_a^3$ in dissipative inflation. \textit{Right}: Comparison between binned and unbinned forecasts, as in Fig.~\ref{fig:bin}. In contrast to the power-law-width cases, the forecast result shows a scaling of $\sigma(f_{\rm NL})\propto H/\gamma$ across most of the parameter space and is less sensitive to binning. This occurs since the folded core has an exponentially narrow width in the weak-dissipation regime, such that the main signals are dominated by the broader support of the template.}
    \label{fig:dissi}
\end{figure}

On the left of Fig.~\ref{fig:dissi}, we show the Fisher forecast constraints as a function of $H/\gamma$. As in the previous examples, the bispectrum provides the dominant source of information. Repeating the forecast with vanishing scale-dependent-bias coefficients $b^{\rm fid}_\zeta=b^{\rm fid}_{\zeta\delta}=0$, we find that the power spectrum constraint changes by a factor of $1.5$, whereas the joint constraint changes by less than $6\%$ at low dissipation. Thus, as for the cutoff-regulated cases, scale-dependent-bias terms mainly affect the power spectrum and are subdominant in the combined constraint. However, marginalization over nuisance parameters considerably weakens the constraint, by up to a factor of $8.5$. This is because the true folded core is exponentially narrow (Eq.~\eqref{eq:dissiwidth}) and not observationally accessible. The majority of the resolvable signal instead comes from the broader support of the template: the logarithmic shoulder near the folded core (see Fig.~\ref{fig:disssquee}), which has
\begin{equation}
    S_{\rm dissi}\sim \frac{\gamma}{H}\log \frac{1}{\tilde{k}_j|\eta_*|},
    \label{eq:shoulder}
\end{equation}
where the $\gamma/H$ factor is due to normalization, together with smoother configurations near the equilateral limit that are comparable in amplitude (see Fig.~\ref{fig:disssquee}). This results in the scaling of $\sigma(f_{\rm NL})\propto H/\gamma$ across the weak-dissipation regime and a much weaker dependence on binning, as shown in Fig.~\ref{fig:dissi}. This indicates that, while the dissipative models can certainly be constrained by the data, we are less sensitive to their particular folded nature.

\subsection{Distinguishability from Standard Templates}
\begin{figure}[!tp]
    \centering
    \includegraphics[width=\linewidth]{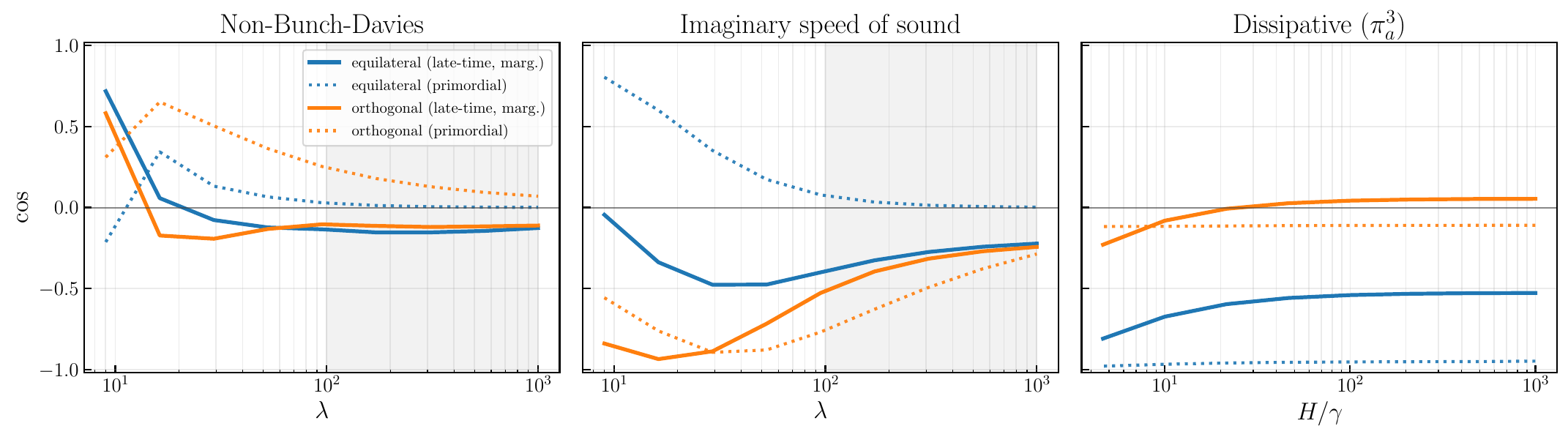}
    \caption{Distinguishability of the folded templates from the standard equilateral and orthogonal templates. Solid curves show the nuisance-marginalized Fisher cosine computed using the binned galaxy bispectrum monopole, while dotted curves show the corresponding primordial shape cosine as a reference. A Fisher cosine close to $\pm 1$ means that the folded signal would be largely captured by the standard templates, while close to $0$ indicates an observationally distinct direction. As $\lambda$ increases, the cutoff-regulated templates exhibit less correlation with the standard templates due to their distinctive localized folded signals. For the dissipation-regulated template, the resolvable signal is dominated by the broader, smoother part of the template, and therefore it has a larger overlap with the equilateral template.}
    \label{fig:cosine}
\end{figure}

Finally, we assess the overlap of the folded-enhanced shape functions with the standard equilateral and orthogonal templates~\cite{Cabass:2022wjy}. This provides a diagnostic of whether the folded signals are captured by existing searches based on the standard templates. We define the late-time cosine between two PNG amplitudes as \citep[e.g.,][]{Cabass:2024wob}
\begin{equation}
    \cos(A,B)=\frac{F_{AB}}{\sqrt{F_{AA}\,F_{BB}}},
\end{equation}
where $F_{AB}$ denotes the Fisher inner product between the bispectrum responses $\partial B_0/\partial f_{\rm NL}^A$ and $\partial B_0/\partial f_{\rm NL}^B$, which is equivalent to a joint forecast of the two PNG shapes. In Fig.~\ref{fig:cosine},
we show the cosine after marginalizing over nuisance parameters, such that the comparison is only between the directions orthogonal to those of the nuisance parameters. For reference, we also show the primordial shape cosines, defined as the inner product between primordial shape functions \citep[e.g.,][]{Babich:2004gb,Senatore:2009gt}. For the power-law-width models, the overlap with both the equilateral and orthogonal templates decreases with increasing $\lambda$. This indicates that the localized enhancement in the folded core is a very distinctive feature of the shapes. Meanwhile, the operator $\pi_a^3$ for dissipative inflation behaves differently. The shape function it induces has a sizable anti-correlation with the equilateral template, mainly coming from the broader support of the template (which can be seen from Fig.~\ref{fig:EIS}). However, the cosine remains significantly different from $-1$ after marginalization and its correlation with the orthogonal template is small. Therefore, a dedicated search for dissipative inflation would still reveal information not captured by standard equilateral and orthogonal constraints. 

\section{Conclusion}
\label{sec5}
\noindent Folded primordial non-Gaussianity provides a well-motivated target for large-scale structure searches beyond the standard local, equilateral, and orthogonal templates~\cite{Holman:2007na, Meerburg:2009ys, Meerburg:2009fi, Aravind:2013lra, Meerburg:2015yka, Greene:2004fln, Schalm:2004qk, Chen:2010bka, Byun:2015rda, Chen:2006nt, Agarwal:2012mq, Albrecht:2014aga,Renaux-Petel:2015mga, Cremonini:2010ua, Brown:2017osf, Mizuno:2017idt, Bjorkmo:2019aev, Christodoulidis:2019mkj, Christodoulidis:2019jsx, Aragam:2019omo, Garcia-Saenz:2025jis, Aoki:2026qea, Garcia-Saenz:2018vqf, Fumagalli:2019noh, Bjorkmo:2019qno, Ferreira:2020qkf, Iarygina:2023msy,LopezNacir:2012rm,Mirbabayi:2022cbt,Green:2020whw,Salcedo:2024smn}. In this work, we have studied the prospects for constraining several representative folded-enhanced bispectra using galaxy clustering. We considered two classes of primordial shape functions with enhancement near the folded regime, $k_1+k_2-k_3\rightarrow0$: (1) cutoff-regulated templates with power-law-width folded cores, represented by non-Bunch-Davies initial states and inflation with an imaginary speed of sound; (2) dissipation-regulated templates with exponentially narrow folded cores, represented by the operator $\pi_a^3$ in the Open EFTofI. We developed a numerical pipeline for computing their contributions to the galaxy power spectrum multipoles and bispectrum monopole within the EFTofLSS framework~\cite{Baumann:2010tm, Ivanov:2021kcd, Carrasco:2012cv}, including the scale-dependent bias induced by their squeezed limits~\cite{Cabass:2018roz}, and used the results to perform Fisher forecasts on the PNG amplitudes.

A common conclusion across the models is that the galaxy bispectrum provides the dominant source of constraining power on folded PNG. This occurs since the folded enhancement is intrinsically a geometric feature of triangle configurations, which is directly accessible to the three-dimensional bispectrum. 
By contrast, the galaxy power spectrum probes the primordial enhancement only indirectly through the scale-dependent bias and loop corrections, which average over the detailed triangle dependence. 

For the cutoff-regulated non-Bunch-Davies and imaginary speed-of-sound templates, the width of the folded core decreases as $1/\lambda$. Here $\lambda$ denotes the large hierarchy controlling the finite-time cutoff: $\lambda=\Lambda_c/H$ for the non-Bunch-Davies template and $\lambda=|m_s|/H$ for the imaginary-speed-of-sound template (see Sec.\,\ref{sec2}). As $\lambda$ increases, the enhanced signal becomes increasingly localized near the exact folded boundary ($k_1+k_2-k_3=0$) and more distinct from smooth gravitational and galaxy-formation contributions. Consequently, marginalization over nuisance parameters produces only a mild loss of information at large $\lambda$, and the folded signals also become less correlated with the standard equilateral and orthogonal templates. This localization, however, makes the constraining power sensitive to Fourier-space  binning. In the idealized unbinned limit, the shrinking folded core gives $\sigma(f_{\rm NL})\propto\sqrt{\lambda}$ (normalizing such that all templates have the same folded-limit amplitude). Once the core width becomes narrower than the bin width, the signal is diluted and the scaling weakens to $\sigma(f_{\rm NL})\propto\lambda$.

The dissipative template exhibits qualitatively different behavior. Assuming weak dissipation, the width of the folded-enhanced core scales as $e^{-H/\gamma}\ll 1$, and is therefore difficult to access with realistic survey volumes. The observable signal is instead dominated by the broader support of the template, including the logarithmic shoulder near the folded core as in Eq.~\eqref{eq:shoulder} and smoother configurations near the equilateral limit (see Fig.~\ref{fig:disssquee}). As a result, the constraining power is much less sensitive to binning, but also has a larger overlap with the equilateral template and galaxy formation signatures. Nevertheless, the dissipative signal is not fully degenerate with equilateral PNG, and a dedicated search would probe information not captured by standard equilateral constraints. 

A major conclusion of this work is that the detectability of folded PNG depends not only on the amplitude of the folded enhancement, but also on its width and morphology. In particular, a universal folded template is unlikely to capture the observational signatures from different inflationary scenarios. We stress that the methods developed herein are not specific to the inflationary models; they can be similarly applied to any folded (or non-folded) primordial shape.

This work can be extended in several directions. On the theory side, one could include the full one-loop galaxy bispectrum (computed for Gaussian initial conditions in~\citep{Philcox:2022frc,DAmico:2022ukl,Bakx:2025pop}). It would also be interesting to explore a systematic IR-resummation of the rapidly oscillatory features in the non-Bunch-Davies template~\cite{Vasudevan:2019ewf}, which is expected to lead to suppressed oscillations on quasi-linear scales. On the observational side, one could apply the same pipeline to a more realistic survey setup, or directly to data from observational surveys, following \citep{Chudaykin:2025vdh}. More broadly, the numerical framework developed here can be applied to other non-separable primordial bispectra, including the full dissipative models, cosmological collider shapes, and more realistic strongly-non-geodesic-motion models, facilitating robust searches for novel inflationary signatures. 

\begin{acknowledgments}
{\footnotesize
\noindent We thank Santiago Agüí-Salcedo, Thomas Bakx, Paolo Creminelli, Mikhail Ivanov, Nickolas Kokron, Mehrdad Mirbabayi, Noah Sailer, Delon Shen, Marko Simonović and Beatriz Tucci for insightful discussions and comments on the draft.}
\end{acknowledgments}

\appendix
\section{Multipole Decomposition}
\label{app:MD}
\noindent In this section, we describe the semi-analytic procedure to facilitate efficient calculations of the redshift-space power spectrum multipoles and bispectrum monopole, which we use to compute the PNG loop contributions in the forecast. Note that, in principle, the analytical treatment is broken by the direction-dependent damping scale $\Sigma^2(\mu)$ used in IR resummation \citep[e.g.,][]{Chudaykin:2020aoj}, though such effects are suppressed by both loops and $f_{\rm NL}$ for the PNG loop contributions.

The power spectrum multipoles are defined as 
\begin{equation}
    P_\ell(k)=(2\ell+1)\int_{-1}^1\frac{d\mu}{2} \,\mathcal{L}_\ell(\mu)P(k,\mu).
\end{equation}
The angular integral is easy to compute for the tree-level contributions to $P(k,\mu)$. For the loop contributions, such as $P_{12}^{\rm NG}$ in Eq.~\eqref{eq:PNG12}, there exists an integral over loop momentum $\bq$:
\begin{equation}
    P^{\rm NG}_{12}(k,\mu)=2Z_1(\bk,\mu)\int\frac{d^3q}{(2\pi)^3}Z_2(\bq,\bk-\bq,\mu)B_{111}(k,q,|\bk-\bq|),
\end{equation}
where the integrand depends only on scalar quantities like $\bq\cdot\bk$, $|\bk|^2$ and $\bk\cdot\bz$. Parametrizing by
\begin{equation}
    q=|\bq|,\qquad
    \eta=\hat{\bq}\cdot\hat{\bk},\qquad
    \zeta=\hat{\bq}\cdot\hat{\bz},
\end{equation}
and the azimuthal angle $\varphi$ of $\bq$ around $\bk$, we can compute part of the multipole decomposition analytically as follows:
\begin{align}
    P_\ell(k)&\sim\int_{-1}^{1}\frac{d\mu}{2}
    Z_1(k,\mu)\int\frac{d^3q}{(2\pi)^3}\mathcal{L}_\ell(\mu)Z_2(q,k,\eta,\mu,\zeta)B(k,q,\eta)\\
    &\rightarrow
    \frac{1}{(2\pi)^2}\int_{-1}^{1}d\eta
    \int q^2dq
    \underbrace{\int_{-1}^{1}\frac{d\mu}{2}\int_0^{2\pi}\frac{d\varphi}{2\pi}\mathcal{L}_\ell(\mu) Z_1(k,\mu)Z_2(q,k,\eta,\mu,\zeta)}_{\rm analytic}B(k,q,\eta),
\end{align}
noting that $Z_2$ is polynomial in $\zeta$, which can be expressed in terms of $\eta$ and $\varphi$ via
\begin{equation}
    \label{eq:zeta}
    \zeta=\mu\eta+\sqrt{1-\mu^2}\sqrt{1-\eta^2}\cos\varphi.
\end{equation}
The $\varphi$-integral can then be carried out using moments of $\cos\varphi$. For example, consider the piece
\begin{equation}
    2Z_1(\bk)Z_2(\bq,\bk-\bq,\mu)\supset 
    2b_1^2\frac{f\mu k}{2}\left(\frac{\mu_\bq}{q}+\frac{\mu_{\bk-\bq}}{|\bk-\bq|}\right)=b_1^2f\mu\frac{(1-2\,r\,\eta)\,\zeta+r\,\mu}{1+r^2-2\,r\,\eta},
\end{equation}
where $r=q/k$. Plugging in Eq.~\eqref{eq:zeta} and integrating over $\mu$ and $\varphi$, we find the multipoles  
\begin{equation}
    \ell=0:\,b_1^2\,f\,\frac{r+\eta-2\,r\,\eta^2}{3\,r\,(1+r^2-2\,r\,\eta)},\qquad     \ell=2:\,b_1^2\,f\,\frac{2\,(r+\eta-2\,r\,\eta^2)}{3\,r\,(1+r^2-2\,r\,\eta)},
\end{equation}
with zero for $\ell=4$. The remainder of the integral can be carried out numerically. 

For the tree-level bispectrum, we define the multipoles,
\begin{equation}
B_\ell(k_1,k_2,k_3)
=
(2\ell+1)\int_{-1}^{1} \frac{d\mu}{2}
\int_{0}^{2\pi} \frac{d\phi}{2\pi}
\,\mathcal{L}_\ell(\mu_1)
B\!\left(
k_1, k_2, k_3,
\mu_1[\mu],
\mu_2[\mu,\phi],
\mu_3[\mu,\phi]
\right),
\end{equation}
where $\mu_1 = \mu, \,\mu_2 = \mu \cos\alpha - \sqrt{1-\mu^2}\,\sin\alpha\,\cos\phi,$
and $\cos\alpha \equiv \hat{\mathbf{k}}_1 \cdot \hat{\mathbf{k}}_2$. Here, there is no loop momentum present, but one still needs to average over the orientation of the triangle relative to the line of sight. This calculation can be performed similarly to that of the power spectrum, noting that $\bk_1+\bk_2+\bk_3=0$, and
\begin{equation}
    \cos\alpha
    =
    \hat{\bk}_1\cdot\hat{\bk}_2
    =
    \frac{k_3^2-k_1^2-k_2^2}{2k_1k_2},\qquad \mu_3=-\frac{k_1\mu_1+k_2\mu_2}{k_3}.
\end{equation}

\bibliography{refs}

@article{Babich:2004gb,
    author = "Babich, Daniel and Creminelli, Paolo and Zaldarriaga, Matias",
    title = "{The Shape of non-Gaussianities}",
    eprint = "astro-ph/0405356",
    archivePrefix = "arXiv",
    reportNumber = "HUTP-04-A022",
    doi = "10.1088/1475-7516/2004/08/009",
    journal = "JCAP",
    volume = "08",
    pages = "009",
    year = "2004"
}

@article{Desjacques:2008vf,
    author = "Desjacques, Vincent and Seljak, Uros and Iliev, Ilian",
    title = "{Scale-dependent bias induced by local non-Gaussianity: A comparison to N-body simulations}",
    eprint = "0811.2748",
    archivePrefix = "arXiv",
    primaryClass = "astro-ph",
    doi = "10.1111/j.1365-2966.2009.14721.x",
    journal = "Mon. Not. Roy. Astron. Soc.",
    volume = "396",
    pages = "85--96",
    year = "2009"
}

@article{Contaldi:2003zv,
    author = "Contaldi, Carlo R. and Peloso, Marco and Kofman, Lev and Linde, Andrei D.",
    title = "{Suppressing the lower multipoles in the CMB anisotropies}",
    eprint = "astro-ph/0303636",
    archivePrefix = "arXiv",
    doi = "10.1088/1475-7516/2003/07/002",
    journal = "JCAP",
    volume = "07",
    pages = "002",
    year = "2003"
}

@article{Holman:2007na,
    author = "Holman, R. and Tolley, Andrew J.",
    title = "{Enhanced Non-Gaussianity from Excited Initial States}",
    eprint = "0710.1302",
    archivePrefix = "arXiv",
    primaryClass = "hep-th",
    reportNumber = "PI-COSMO-64",
    doi = "10.1088/1475-7516/2008/05/001",
    journal = "JCAP",
    volume = "05",
    pages = "001",
    year = "2008"
}

@article{Meerburg:2009ys,
    author = "Meerburg, Pieter Daniel and van der Schaar, Jan Pieter and Corasaniti, Pier Stefano",
    title = "{Signatures of Initial State Modifications on Bispectrum Statistics}",
    eprint = "0901.4044",
    archivePrefix = "arXiv",
    primaryClass = "hep-th",
    reportNumber = "ITFA-2008-53",
    doi = "10.1088/1475-7516/2009/05/018",
    journal = "JCAP",
    volume = "05",
    pages = "018",
    year = "2009"
}

@article{Meerburg:2009fi,
    author = "Meerburg, P. Daniel and van der Schaar, Jan Pieter and Jackson, Mark G.",
    title = "{Bispectrum signatures of a modified vacuum in single field inflation with a small speed of sound}",
    eprint = "0910.4986",
    archivePrefix = "arXiv",
    primaryClass = "hep-th",
    doi = "10.1088/1475-7516/2010/02/001",
    journal = "JCAP",
    volume = "02",
    pages = "001",
    year = "2010"
}

@article{Aravind:2013lra,
    author = "Aravind, Aditya and Lorshbough, Dustin and Paban, Sonia",
    title = "{Non-Gaussianity from Excited Initial Inflationary States}",
    eprint = "1303.1440",
    archivePrefix = "arXiv",
    primaryClass = "hep-th",
    reportNumber = "UTTG-05-13, TCC-004-13",
    doi = "10.1007/JHEP07(2013)076",
    journal = "JHEP",
    volume = "07",
    pages = "076",
    year = "2013"
}

@article{Salcedo:2026sdn,
    author = {Salcedo, Santiago Ag{\"u}{\'\i} and Colas, Thomas and Suman, Petar and Zhang, Bowei and Fergusson, James and Shellard, E. P. S.},
    title = "{Primordial non-Gaussianity constraints on dissipative inflation}",
    eprint = "2603.13473",
    archivePrefix = "arXiv",
    primaryClass = "astro-ph.CO",
    month = "3",
    journal = "",
    year = "2026"
}

@article{Dalal:2007cu,
    author = "Dalal, Neal and Dore, Olivier and Huterer, Dragan and Shirokov, Alexander",
    title = "{The imprints of primordial non-gaussianities on large-scale structure: scale dependent bias and abundance of virialized objects}",
    eprint = "0710.4560",
    archivePrefix = "arXiv",
    primaryClass = "astro-ph",
    doi = "10.1103/PhysRevD.77.123514",
    journal = "Phys. Rev. D",
    volume = "77",
    pages = "123514",
    year = "2008"
}

@article{Assassi:2015fma,
    author = "Assassi, Valentin and Baumann, Daniel and Schmidt, Fabian",
    title = "{Galaxy Bias and Primordial Non-Gaussianity}",
    eprint = "1510.03723",
    archivePrefix = "arXiv",
    primaryClass = "astro-ph.CO",
    doi = "10.1088/1475-7516/2015/12/043",
    journal = "JCAP",
    volume = "12",
    pages = "043",
    year = "2015"
}

@article{Desjacques:2016bnm,
    author = "Desjacques, Vincent and Jeong, Donghui and Schmidt, Fabian",
    title = "{Large-Scale Galaxy Bias}",
    eprint = "1611.09787",
    archivePrefix = "arXiv",
    primaryClass = "astro-ph.CO",
    doi = "10.1016/j.physrep.2017.12.002",
    journal = "Phys. Rept.",
    volume = "733",
    pages = "1--193",
    year = "2018"
}

@article{Cabass:2022ymb,
    author = "Cabass, Giovanni and Ivanov, Mikhail M. and Philcox, Oliver H. E. and Simonovi{\'c}, Marko and Zaldarriaga, Matias",
    title = "{Constraints on multifield inflation from the BOSS galaxy survey}",
    eprint = "2204.01781",
    archivePrefix = "arXiv",
    primaryClass = "astro-ph.CO",
    reportNumber = "CERN-TH-2022-055",
    doi = "10.1103/PhysRevD.106.043506",
    journal = "Phys. Rev. D",
    volume = "106",
    number = "4",
    pages = "043506",
    year = "2022"
}

@article{Ivanov:2023qzb,
    author = "Ivanov, Mikhail M. and Philcox, Oliver H. E. and Cabass, Giovanni and Nishimichi, Takahiro and Simonovi{\'c}, Marko and Zaldarriaga, Matias",
    title = "{Cosmology with the galaxy bispectrum multipoles: Optimal estimation and application to BOSS data}",
    eprint = "2302.04414",
    archivePrefix = "arXiv",
    primaryClass = "astro-ph.CO",
    reportNumber = "YITP-23-13, CERN-TH-2023-022",
    doi = "10.1103/PhysRevD.107.083515",
    journal = "Phys. Rev. D",
    volume = "107",
    number = "8",
    pages = "083515",
    year = "2023"
}

@article{Jiang:2015hfa,
    author = "Jiang, Hongliang and Wang, Yi",
    title = "{Towards the physical vacuum of cosmic inflation}",
    eprint = "1507.05193",
    archivePrefix = "arXiv",
    primaryClass = "hep-th",
    doi = "10.1016/j.physletb.2016.06.069",
    journal = "Phys. Lett. B",
    volume = "760",
    pages = "202--206",
    year = "2016"
}

@article{Green:2020whw,
    author = "Green, Daniel and Porto, Rafael A.",
    title = "{Signals of a Quantum Universe}",
    eprint = "2001.09149",
    archivePrefix = "arXiv",
    primaryClass = "hep-th",
    doi = "10.1103/PhysRevLett.124.251302",
    journal = "Phys. Rev. Lett.",
    volume = "124",
    number = "25",
    pages = "251302",
    year = "2020"
}

@article{Flauger:2013hra,
    author = "Flauger, Raphael and Green, Daniel and Porto, Rafael A.",
    title = "{On squeezed limits in single-field inflation.  Part I}",
    eprint = "1303.1430",
    archivePrefix = "arXiv",
    primaryClass = "hep-th",
    doi = "10.1088/1475-7516/2013/08/032",
    journal = "JCAP",
    volume = "08",
    pages = "032",
    year = "2013"
}

@article{Cabass:2018roz,
    author = "Cabass, Giovanni and Pajer, Enrico and Schmidt, Fabian",
    title = "{Imprints of Oscillatory Bispectra on Galaxy Clustering}",
    eprint = "1804.07295",
    archivePrefix = "arXiv",
    primaryClass = "astro-ph.CO",
    doi = "10.1088/1475-7516/2018/09/003",
    journal = "JCAP",
    volume = "09",
    pages = "003",
    year = "2018"
}

@article{Meerburg:2015yka,
    author = {Meerburg, P. Daniel and M{\"u}nchmeyer, Moritz},
    title = "{Optimal CMB estimators for bispectra from excited states}",
    eprint = "1505.05882",
    archivePrefix = "arXiv",
    primaryClass = "astro-ph.CO",
    doi = "10.1103/PhysRevD.92.063527",
    journal = "Phys. Rev. D",
    volume = "92",
    number = "6",
    pages = "063527",
    year = "2015"
}

@article{Creminelli:2006xe,
    author = "Creminelli, Paolo and Luty, Markus A. and Nicolis, Alberto and Senatore, Leonardo",
    title = "{Starting the Universe: Stable Violation of the Null Energy Condition and Non-standard Cosmologies}",
    eprint = "hep-th/0606090",
    archivePrefix = "arXiv",
    reportNumber = "HUTP-06-A0019, MIT-CTP-3738, IC-2006-034",
    doi = "10.1088/1126-6708/2006/12/080",
    journal = "JHEP",
    volume = "12",
    pages = "080",
    year = "2006"
}

@article{Cheung:2007st,
    author = "Cheung, Clifford and Creminelli, Paolo and Fitzpatrick, A. Liam and Kaplan, Jared and Senatore, Leonardo",
    title = "{The Effective Field Theory of Inflation}",
    eprint = "0709.0293",
    archivePrefix = "arXiv",
    primaryClass = "hep-th",
    reportNumber = "IC-2007-032",
    doi = "10.1088/1126-6708/2008/03/014",
    journal = "JHEP",
    volume = "03",
    pages = "014",
    year = "2008"
}

@article{Renaux-Petel:2015mga,
    author = "Renaux-Petel, S{\'e}bastien and Turzy{\'n}ski, Krzysztof",
    title = "{Geometrical Destabilization of Inflation}",
    eprint = "1510.01281",
    archivePrefix = "arXiv",
    primaryClass = "astro-ph.CO",
    doi = "10.1103/PhysRevLett.117.141301",
    journal = "Phys. Rev. Lett.",
    volume = "117",
    number = "14",
    pages = "141301",
    year = "2016"
}

@article{Bernardeau:2001qr,
    author = "Bernardeau, F. and Colombi, S. and Gaztanaga, E. and Scoccimarro, R.",
    title = "{Large scale structure of the universe and cosmological perturbation theory}",
    eprint = "astro-ph/0112551",
    archivePrefix = "arXiv",
    reportNumber = "SACLAY-T01-142",
    doi = "10.1016/S0370-1573(02)00135-7",
    journal = "Phys. Rept.",
    volume = "367",
    pages = "1--248",
    year = "2002"
}

@article{Weinberg:2005vy,
    author = "Weinberg, Steven",
    title = "{Quantum contributions to cosmological correlations}",
    eprint = "hep-th/0506236",
    archivePrefix = "arXiv",
    reportNumber = "UTTG-01-05",
    doi = "10.1103/PhysRevD.72.043514",
    journal = "Phys. Rev. D",
    volume = "72",
    pages = "043514",
    year = "2005"
}

@article{Maldacena:2002vr,
    author = "Maldacena, Juan Martin",
    title = "{Non-Gaussian features of primordial fluctuations in single field inflationary models}",
    eprint = "astro-ph/0210603",
    archivePrefix = "arXiv",
    doi = "10.1088/1126-6708/2003/05/013",
    journal = "JHEP",
    volume = "05",
    pages = "013",
    year = "2003"
}

@article{Berera:1995ie,
    author = "Berera, Arjun",
    title = "{Warm inflation}",
    eprint = "astro-ph/9509049",
    archivePrefix = "arXiv",
    reportNumber = "PSU-TH-159",
    doi = "10.1103/PhysRevLett.75.3218",
    journal = "Phys. Rev. Lett.",
    volume = "75",
    pages = "3218--3221",
    year = "1995"
}

@article{Salcedo:2024smn,
    author = "Salcedo, Santiago Agui and Colas, Thomas and Pajer, Enrico",
    title = "{The open effective field theory of inflation}",
    eprint = "2404.15416",
    archivePrefix = "arXiv",
    primaryClass = "hep-th",
    doi = "10.1007/JHEP10(2024)248",
    journal = "JHEP",
    volume = "10",
    pages = "248",
    year = "2024"
}

@article{Blas:2016sfa,
    author = "Blas, Diego and Garny, Mathias and Ivanov, Mikhail M. and Sibiryakov, Sergey",
    title = "{Time-Sliced Perturbation Theory II: Baryon Acoustic Oscillations and Infrared Resummation}",
    eprint = "1605.02149",
    archivePrefix = "arXiv",
    primaryClass = "astro-ph.CO",
    reportNumber = "CERN-TH-2016-059, INR-TH-2016-009",
    doi = "10.1088/1475-7516/2016/07/028",
    journal = "JCAP",
    volume = "07",
    pages = "028",
    year = "2016"
}

@article{Oddo:2019run,
    author = "Oddo, Andrea and Sefusatti, Emiliano and Porciani, Cristiano and Monaco, Pierluigi and S{\'a}nchez, Ariel G.",
    title = "{Toward a robust inference method for the galaxy bispectrum: likelihood function and model selection}",
    eprint = "1908.01774",
    archivePrefix = "arXiv",
    primaryClass = "astro-ph.CO",
    doi = "10.1088/1475-7516/2020/03/056",
    journal = "JCAP",
    volume = "03",
    pages = "056",
    year = "2020"
}

@article{MoradinezhadDizgah:2012aya,
    author = "Moradinezhad Dizgah, Azadeh and Gnedin, Nickolay Y. and Kinney, William H.",
    title = "{Reionization History and CMB Parameter Estimation}",
    eprint = "1211.7007",
    archivePrefix = "arXiv",
    primaryClass = "astro-ph.CO",
    reportNumber = "FERMILAB-PUB-12-926-A",
    doi = "10.1088/1475-7516/2013/05/017",
    journal = "JCAP",
    volume = "05",
    pages = "017",
    year = "2013"
}

@article{MoradinezhadDizgah:2020whw,
    author = "Moradinezhad Dizgah, Azadeh and Biagetti, Matteo and Sefusatti, Emiliano and Desjacques, Vincent and Nore{\~n}a, Jorge",
    title = "{Primordial Non-Gaussianity from Biased Tracers: Likelihood Analysis of Real-Space Power Spectrum and Bispectrum}",
    eprint = "2010.14523",
    archivePrefix = "arXiv",
    primaryClass = "astro-ph.CO",
    doi = "10.1088/1475-7516/2021/05/015",
    journal = "JCAP",
    volume = "05",
    pages = "015",
    year = "2021"
}

@article{Chen:2024bdg,
    author = "Chen, Shu-Fan and Chakraborty, Priyesh and Dvorkin, Cora",
    title = "{Analysis of BOSS galaxy data with weighted skew-spectra}",
    eprint = "2401.13036",
    archivePrefix = "arXiv",
    primaryClass = "astro-ph.CO",
    doi = "10.1088/1475-7516/2024/05/011",
    journal = "JCAP",
    volume = "05",
    pages = "011",
    year = "2024"
}

@article{Baumann:2021ykm,
    author = "Baumann, Daniel and Green, Daniel",
    title = "{The power of locality: primordial non-Gaussianity at the map level}",
    eprint = "2112.14645",
    archivePrefix = "arXiv",
    primaryClass = "astro-ph.CO",
    doi = "10.1088/1475-7516/2022/08/061",
    journal = "JCAP",
    volume = "08",
    number = "08",
    pages = "061",
    year = "2022"
}

@article{DESI:2024aax,
    author = "Adame, A. G. and others",
    collaboration = "DESI",
    title = "{DESI 2024 II: sample definitions, characteristics, and two-point clustering statistics}",
    eprint = "2411.12020",
    archivePrefix = "arXiv",
    primaryClass = "astro-ph.CO",
    reportNumber = "FERMILAB-PUB-24-0850-PPD",
    doi = "10.1088/1475-7516/2025/07/017",
    journal = "JCAP",
    volume = "07",
    pages = "017",
    year = "2025"
}

@article{Ivanov:2018gjr,
    author = "Ivanov, Mikhail M. and Sibiryakov, Sergey",
    title = "{Infrared Resummation for Biased Tracers in Redshift Space}",
    eprint = "1804.05080",
    archivePrefix = "arXiv",
    primaryClass = "astro-ph.CO",
    reportNumber = "CERN-TH-2018-076, INR-TH-2018-006",
    doi = "10.1088/1475-7516/2018/07/053",
    journal = "JCAP",
    volume = "07",
    pages = "053",
    year = "2018"
}

@article{Vasudevan:2019ewf,
    author = "Vasudevan, Anagha and Ivanov, Mikhail M. and Sibiryakov, Sergey and Lesgourgues, Julien",
    title = "{Time-sliced perturbation theory with primordial non-Gaussianity and effects of large bulk flows on inflationary oscillating features}",
    eprint = "1906.08697",
    archivePrefix = "arXiv",
    primaryClass = "astro-ph.CO",
    reportNumber = "CERN-TH-2019-091, INR-TH-2019-012, TTK-19-22",
    doi = "10.1088/1475-7516/2019/09/037",
    journal = "JCAP",
    volume = "09",
    pages = "037",
    year = "2019"
}

@article{Assassi:2015jqa,
    author = "Assassi, Valentin and Baumann, Daniel and Pajer, Enrico and Welling, Yvette and van der Woude, Drian",
    title = "{Effective theory of large-scale structure with primordial non-Gaussianity}",
    eprint = "1505.06668",
    archivePrefix = "arXiv",
    primaryClass = "astro-ph.CO",
    doi = "10.1088/1475-7516/2015/11/024",
    journal = "JCAP",
    volume = "11",
    pages = "024",
    year = "2015"
}

@article{Cabass:2022wjy,
    author = "Cabass, Giovanni and Ivanov, Mikhail M. and Philcox, Oliver H. E. and Simonovi{\'c}, Marko and Zaldarriaga, Matias",
    title = "{Constraints on Single-Field Inflation from the BOSS Galaxy Survey}",
    eprint = "2201.07238",
    archivePrefix = "arXiv",
    primaryClass = "astro-ph.CO",
    reportNumber = "CERN-TH-2022-005",
    doi = "10.1103/PhysRevLett.129.021301",
    journal = "Phys. Rev. Lett.",
    volume = "129",
    number = "2",
    pages = "021301",
    year = "2022"
}

@article{Ivanov:2021kcd,
    author = "Ivanov, Mikhail M. and Philcox, Oliver H. E. and Nishimichi, Takahiro and Simonovi{\'c}, Marko and Takada, Masahiro and Zaldarriaga, Matias",
    title = "{Precision analysis of the redshift-space galaxy bispectrum}",
    eprint = "2110.10161",
    archivePrefix = "arXiv",
    primaryClass = "astro-ph.CO",
    reportNumber = "YITP-21-120, CERN-TH-2021-155",
    doi = "10.1103/PhysRevD.105.063512",
    journal = "Phys. Rev. D",
    volume = "105",
    number = "6",
    pages = "063512",
    year = "2022"
}

@article{Cabass:2022epm,
    author = "Cabass, Giovanni and Ivanov, Mikhail M. and Philcox, Oliver H. E. and Simonovic, Marko and Zaldarriaga, Matias",
    title = "{Constraining single-field inflation with MegaMapper}",
    eprint = "2211.14899",
    archivePrefix = "arXiv",
    primaryClass = "astro-ph.CO",
    reportNumber = "CERN-TH-2022-203",
    doi = "10.1016/j.physletb.2023.137912",
    journal = "Phys. Lett. B",
    volume = "841",
    pages = "137912",
    year = "2023"
}

@article{Greene:2004fln,
    author = "Greene, Brian and Schalm, Koenraad and van der Schaar, Jan Pieter and Shiu, Gary",
    editor = "Chen, P. and Bloom, Elliott D. and Madejski, G. and Petrosian, V.",
    title = "{Extracting new physics from the CMB}",
    eprint = "astro-ph/0503458",
    archivePrefix = "arXiv",
    reportNumber = "TSRA-2004-0001",
    journal = "eConf",
    volume = "C041213",
    pages = "0001",
    year = "2004"
}

@article{Schalm:2004qk,
    author = "Schalm, Koenraad and Shiu, Gary and van der Schaar, Jan Pieter",
    title = "{Decoupling in an expanding universe: Boundary RG flow affects initial conditions for inflation}",
    eprint = "hep-th/0401164",
    archivePrefix = "arXiv",
    reportNumber = "CERN-PH-TH-2004-001, CU-TP-1103, MAD-TH-03-7",
    doi = "10.1088/1126-6708/2004/04/076",
    journal = "JHEP",
    volume = "04",
    pages = "076",
    year = "2004"
}

@article{Matarrese:2008nc,
    author = "Matarrese, Sabino and Verde, Licia",
    title = "{The effect of primordial non-Gaussianity on halo bias}",
    eprint = "0801.4826",
    archivePrefix = "arXiv",
    primaryClass = "astro-ph",
    doi = "10.1086/587840",
    journal = "Astrophys. J. Lett.",
    volume = "677",
    pages = "L77--L80",
    year = "2008"
}

@article{Baldauf:2015vio,
    author = "Baldauf, Tobias and Seljak, Uro{\v{s}} and Senatore, Leonardo and Zaldarriaga, Matias",
    title = "{Linear response to long wavelength fluctuations using curvature simulations}",
    eprint = "1511.01465",
    archivePrefix = "arXiv",
    primaryClass = "astro-ph.CO",
    doi = "10.1088/1475-7516/2016/09/007",
    journal = "JCAP",
    volume = "09",
    pages = "007",
    year = "2016"
}

@article{Schmidt:2010gw,
    author = "Schmidt, Fabian and Kamionkowski, Marc",
    title = "{Halo Clustering with Non-Local Non-Gaussianity}",
    eprint = "1008.0638",
    archivePrefix = "arXiv",
    primaryClass = "astro-ph.CO",
    doi = "10.1103/PhysRevD.82.103002",
    journal = "Phys. Rev. D",
    volume = "82",
    pages = "103002",
    year = "2010"
}

@article{Scoccimarro:2015bla,
    author = "Scoccimarro, Roman",
    title = "{Fast Estimators for Redshift-Space Clustering}",
    eprint = "1506.02729",
    archivePrefix = "arXiv",
    primaryClass = "astro-ph.CO",
    doi = "10.1103/PhysRevD.92.083532",
    journal = "Phys. Rev. D",
    volume = "92",
    number = "8",
    pages = "083532",
    year = "2015"
}

@article{Chudaykin:2020aoj,
    author = "Chudaykin, Anton and Ivanov, Mikhail M. and Philcox, Oliver H. E. and Simonovi{\'c}, Marko",
    title = "{Nonlinear perturbation theory extension of the Boltzmann code CLASS}",
    eprint = "2004.10607",
    archivePrefix = "arXiv",
    primaryClass = "astro-ph.CO",
    reportNumber = "INR-TH-2020-016, CERN-TH-2020-062",
    doi = "10.1103/PhysRevD.102.063533",
    journal = "Phys. Rev. D",
    volume = "102",
    number = "6",
    pages = "063533",
    year = "2020"
}

@article{Bakx:2025pop,
    author = "Bakx, Thomas and Ivanov, Mikhail M. and Philcox, Oliver H. E. and Vlah, Zvonimir",
    title = "{One-Loop Galaxy Bispectrum: Consistent Theory, Efficient Analysis with COBRA, and Implications for Cosmological Parameters}",
    eprint = "2507.22110",
    archivePrefix = "arXiv",
    primaryClass = "astro-ph.CO",
    reportNumber = "MIT-CTP/5891, RBI-ThPhys-2025-23",
    month = "7",
    year = "2025",
    journal = ""
}

@article{Chudaykin:2019ock,
    author = "Chudaykin, Anton and Ivanov, Mikhail M.",
    title = "{Measuring neutrino masses with large-scale structure: Euclid forecast with controlled theoretical error}",
    eprint = "1907.06666",
    archivePrefix = "arXiv",
    primaryClass = "astro-ph.CO",
    reportNumber = "INR-TH-2019-014",
    doi = "10.1088/1475-7516/2019/11/034",
    journal = "JCAP",
    volume = "11",
    pages = "034",
    year = "2019"
}

@article{Chen:2005fe,
    author = "Chen, Xingang",
    title = "{Running non-Gaussianities in DBI inflation}",
    eprint = "astro-ph/0507053",
    archivePrefix = "arXiv",
    reportNumber = "UFIFT-HEP-05-12",
    doi = "10.1103/PhysRevD.72.123518",
    journal = "Phys. Rev. D",
    volume = "72",
    pages = "123518",
    year = "2005"
}

@article{LopezNacir:2012rm,
    author = "Lopez Nacir, Diana and Porto, Rafael A. and Zaldarriaga, Matias",
    title = "{The consistency condition for the three-point function in dissipative single-clock inflation}",
    eprint = "1206.7083",
    archivePrefix = "arXiv",
    primaryClass = "hep-th",
    doi = "10.1088/1475-7516/2012/09/004",
    journal = "JCAP",
    volume = "09",
    pages = "004",
    year = "2012"
}

@article{Cheung:2007sv,
    author = "Cheung, Clifford and Fitzpatrick, A. Liam and Kaplan, Jared and Senatore, Leonardo",
    title = "{On the consistency relation of the 3-point function in single field inflation}",
    eprint = "0709.0295",
    archivePrefix = "arXiv",
    primaryClass = "hep-th",
    doi = "10.1088/1475-7516/2008/02/021",
    journal = "JCAP",
    volume = "02",
    pages = "021",
    year = "2008"
}

@article{Fumagalli:2019noh,
    author = "Fumagalli, Jacopo and Garcia-Saenz, Sebastian and Pinol, Lucas and Renaux-Petel, S{\'e}bastien and Ronayne, John",
    title = "{Hyper-Non-Gaussianities in Inflation with Strongly Nongeodesic Motion}",
    eprint = "1902.03221",
    archivePrefix = "arXiv",
    primaryClass = "hep-th",
    doi = "10.1103/PhysRevLett.123.201302",
    journal = "Phys. Rev. Lett.",
    volume = "123",
    number = "20",
    pages = "201302",
    year = "2019"
}

@article{Garcia-Saenz:2025jis,
    author = "Garcia-Saenz, Sebastian and Lu, Yizhou and Renaux-Petel, S{\'e}bastien",
    title = "{Loops in inflation with strongly non-geodesic motion}",
    eprint = "2503.18516",
    archivePrefix = "arXiv",
    primaryClass = "hep-th",
    doi = "10.1088/1475-7516/2025/11/039",
    journal = "JCAP",
    volume = "11",
    pages = "039",
    year = "2025"
}

@article{Aoki:2026qea,
    author = "Aoki, Shuntaro and Roest, Diederik and Werth, Denis",
    title = "{Universal Non-Gaussian Signatures from Transient Instabilities}",
    eprint = "2604.01035",
    archivePrefix = "arXiv",
    primaryClass = "astro-ph.CO",
    reportNumber = "RIKEN-iTHEMS-Report-26",
    journal = "",
    month = "4",
    year = "2026"
}

@article{Cremonini:2010ua,
    author = "Cremonini, Sera and Lalak, Zygmunt and Turzynski, Krzysztof",
    title = "{Strongly Coupled Perturbations in Two-Field Inflationary Models}",
    eprint = "1010.3021",
    archivePrefix = "arXiv",
    primaryClass = "hep-th",
    reportNumber = "IFT-10-14, DAMTP-2010-80, MIFPA-10-43",
    doi = "10.1088/1475-7516/2011/03/016",
    journal = "JCAP",
    volume = "03",
    pages = "016",
    year = "2011"
}

@article{Brown:2017osf,
    author = "Brown, Adam R.",
    title = "{Hyperbolic Inflation}",
    eprint = "1705.03023",
    archivePrefix = "arXiv",
    primaryClass = "hep-th",
    doi = "10.1103/PhysRevLett.121.251601",
    journal = "Phys. Rev. Lett.",
    volume = "121",
    number = "25",
    pages = "251601",
    year = "2018"
}

@article{Mizuno:2017idt,
    author = "Mizuno, Shuntaro and Mukohyama, Shinji",
    title = "{Primordial perturbations from inflation with a hyperbolic field-space}",
    eprint = "1707.05125",
    archivePrefix = "arXiv",
    primaryClass = "hep-th",
    reportNumber = "YITP-17-73, IPMU17-0099",
    doi = "10.1103/PhysRevD.96.103533",
    journal = "Phys. Rev. D",
    volume = "96",
    number = "10",
    pages = "103533",
    year = "2017"
}

@article{Bjorkmo:2019aev,
    author = "Bjorkmo, Theodor and Marsh, M. C. David",
    title = "{Hyperinflation generalised: from its attractor mechanism to its tension with the {\textquoteleft}swampland conditions{\textquoteright}}",
    eprint = "1901.08603",
    archivePrefix = "arXiv",
    primaryClass = "hep-th",
    doi = "10.1007/JHEP04(2019)172",
    journal = "JHEP",
    volume = "04",
    pages = "172",
    year = "2019"
}

@article{Christodoulidis:2019mkj,
    author = "Christodoulidis, Perseas and Roest, Diederik and Sfakianakis, Evangelos I.",
    title = "{Attractors, Bifurcations and Curvature in Multi-field Inflation}",
    eprint = "1903.03513",
    archivePrefix = "arXiv",
    primaryClass = "gr-qc",
    reportNumber = "Nikhef 2019-008",
    doi = "10.1088/1475-7516/2020/08/006",
    journal = "JCAP",
    volume = "08",
    pages = "006",
    year = "2020"
}

@article{Christodoulidis:2019jsx,
    author = "Christodoulidis, Perseas and Roest, Diederik and Sfakianakis, Evangelos I.",
    title = "{Scaling attractors in multi-field inflation}",
    eprint = "1903.06116",
    archivePrefix = "arXiv",
    primaryClass = "hep-th",
    reportNumber = "Nikhef 2019-007",
    doi = "10.1088/1475-7516/2019/12/059",
    journal = "JCAP",
    volume = "12",
    pages = "059",
    year = "2019"
}

@article{Garcia-Saenz:2018vqf,
    author = "Garcia-Saenz, Sebastian and Renaux-Petel, S{\'e}bastien",
    title = "{Flattened non-Gaussianities from the effective field theory of inflation with imaginary speed of sound}",
    eprint = "1805.12563",
    archivePrefix = "arXiv",
    primaryClass = "hep-th",
    doi = "10.1088/1475-7516/2018/11/005",
    journal = "JCAP",
    volume = "11",
    pages = "005",
    year = "2018"
}

@article{Bjorkmo:2019qno,
    author = "Bjorkmo, Theodor and Ferreira, Ricardo Z. and Marsh, M. C. David",
    title = "{Mild Non-Gaussianities under Perturbative Control from Rapid-Turn Inflation Models}",
    eprint = "1908.11316",
    archivePrefix = "arXiv",
    primaryClass = "hep-th",
    doi = "10.1088/1475-7516/2019/12/036",
    journal = "JCAP",
    volume = "12",
    pages = "036",
    year = "2019"
}

@article{Ferreira:2020qkf,
    author = "Ferreira, Ricardo Z.",
    title = "{Non-Gaussianities in models of inflation with large and negative entropic masses}",
    eprint = "2003.13410",
    archivePrefix = "arXiv",
    primaryClass = "astro-ph.CO",
    doi = "10.1088/1475-7516/2020/08/034",
    journal = "JCAP",
    volume = "08",
    pages = "034",
    year = "2020"
}

@article{Iarygina:2023msy,
    author = "Iarygina, Oksana and Marsh, M. C. David and Salinas, Gustavo",
    title = "{Non-Gaussianity in rapid-turn multi-field inflation}",
    eprint = "2303.14156",
    archivePrefix = "arXiv",
    primaryClass = "astro-ph.CO",
    reportNumber = "NORDITA 2023-015",
    doi = "10.1088/1475-7516/2024/03/014",
    journal = "JCAP",
    volume = "03",
    pages = "014",
    year = "2024"
}

@article{Aragam:2019omo,
    author = "Aragam, Vikas and Paban, Sonia and Rosati, Robert",
    title = "{Multi-field Inflation in High-Slope Potentials}",
    eprint = "1905.07495",
    archivePrefix = "arXiv",
    primaryClass = "hep-th",
    reportNumber = "UTTG-05-19",
    doi = "10.1088/1475-7516/2020/04/022",
    journal = "JCAP",
    volume = "04",
    pages = "022",
    year = "2020"
}

@article{Chen:2010bka,
    author = "Chen, Xingang",
    title = "{Folded Resonant Non-Gaussianity in General Single Field Inflation}",
    eprint = "1008.2485",
    archivePrefix = "arXiv",
    primaryClass = "hep-th",
    doi = "10.1088/1475-7516/2010/12/003",
    journal = "JCAP",
    volume = "12",
    pages = "003",
    year = "2010"
}

@article{Byun:2015rda,
    author = "Byun, Joyce and Agarwal, Nishant and Bean, Rachel and Holman, Richard",
    title = "{Looking for non-Gaussianity in all the right places: A new basis for nonseparable bispectra}",
    eprint = "1504.01394",
    archivePrefix = "arXiv",
    primaryClass = "astro-ph.CO",
    doi = "10.1103/PhysRevD.91.123518",
    journal = "Phys. Rev. D",
    volume = "91",
    number = "12",
    pages = "123518",
    year = "2015"
}

@article{Mirbabayi:2022cbt,
    author = "Mirbabayi, Mehrdad and Gruzinov, Andrei",
    title = "{Shapes of non-Gaussianity in warm inflation}",
    eprint = "2205.13227",
    archivePrefix = "arXiv",
    primaryClass = "astro-ph.CO",
    doi = "10.1088/1475-7516/2023/02/012",
    journal = "JCAP",
    volume = "02",
    pages = "012",
    year = "2023"
}

@article{Chen:2006nt,
    author = "Chen, Xingang and Huang, Min-xin and Kachru, Shamit and Shiu, Gary",
    title = "{Observational signatures and non-Gaussianities of general single field inflation}",
    eprint = "hep-th/0605045",
    archivePrefix = "arXiv",
    reportNumber = "SLAC-PUB-11840, MAD-TH-06-3, UFIFT-HEP-06-9, SU-ITP-06-12, CU-TP-1147",
    doi = "10.1088/1475-7516/2007/01/002",
    journal = "JCAP",
    volume = "01",
    pages = "002",
    year = "2007"
}

@article{Agarwal:2012mq,
    author = "Agarwal, Nishant and Holman, R. and Tolley, Andrew J. and Lin, Jennifer",
    title = "{Effective field theory and non-Gaussianity from general inflationary states}",
    eprint = "1212.1172",
    archivePrefix = "arXiv",
    primaryClass = "hep-th",
    doi = "10.1007/JHEP05(2013)085",
    journal = "JHEP",
    volume = "05",
    pages = "085",
    year = "2013"
}

@article{Albrecht:2014aga,
    author = "Albrecht, Andreas and Bolis, Nadia and Holman, R.",
    title = "{Cosmological Consequences of Initial State Entanglement}",
    eprint = "1408.6859",
    archivePrefix = "arXiv",
    primaryClass = "hep-th",
    doi = "10.1007/JHEP11(2014)093",
    journal = "JHEP",
    volume = "11",
    pages = "093",
    year = "2014"
}

@article{Lyth:1996im,
    author = "Lyth, David H.",
    title = "{What would we learn by detecting a gravitational wave signal in the cosmic microwave background anisotropy?}",
    eprint = "hep-ph/9606387",
    archivePrefix = "arXiv",
    reportNumber = "LANCASTER-TH-9612",
    doi = "10.1103/PhysRevLett.78.1861",
    journal = "Phys. Rev. Lett.",
    volume = "78",
    pages = "1861--1863",
    year = "1997"
}

@article{BICEP:2021xfz,
    author = "Ade, P. A. R. and others",
    collaboration = "BICEP, Keck",
    title = "{Improved Constraints on Primordial Gravitational Waves using Planck, WMAP, and BICEP/Keck Observations through the 2018 Observing Season}",
    eprint = "2110.00483",
    archivePrefix = "arXiv",
    primaryClass = "astro-ph.CO",
    doi = "10.1103/PhysRevLett.127.151301",
    journal = "Phys. Rev. Lett.",
    volume = "127",
    number = "15",
    pages = "151301",
    year = "2021"
}

@article{Bartolo:2004if,
    author = "Bartolo, N. and Komatsu, E. and Matarrese, Sabino and Riotto, A.",
    title = "{Non-Gaussianity from inflation: Theory and observations}",
    eprint = "astro-ph/0406398",
    archivePrefix = "arXiv",
    reportNumber = "DFPD-04-A-12",
    doi = "10.1016/j.physrep.2004.08.022",
    journal = "Phys. Rept.",
    volume = "402",
    pages = "103--266",
    year = "2004"
}

@article{Cabass:2016cgp,
    author = "Cabass, Giovanni and Pajer, Enrico and Schmidt, Fabian",
    title = "{How Gaussian can our Universe be?}",
    eprint = "1612.00033",
    archivePrefix = "arXiv",
    primaryClass = "hep-th",
    doi = "10.1088/1475-7516/2017/01/003",
    journal = "JCAP",
    volume = "01",
    pages = "003",
    year = "2017"
}

@article{Enqvist:2001zp,
    author = "Enqvist, Kari and Sloth, Martin S.",
    title = "{Adiabatic CMB perturbations in pre - big bang string cosmology}",
    eprint = "hep-ph/0109214",
    archivePrefix = "arXiv",
    reportNumber = "HIP-2001-51-TH",
    doi = "10.1016/S0550-3213(02)00043-3",
    journal = "Nucl. Phys. B",
    volume = "626",
    pages = "395--409",
    year = "2002"
}

@article{Lyth:2001nq,
    author = "Lyth, David H. and Wands, David",
    title = "{Generating the curvature perturbation without an inflaton}",
    eprint = "hep-ph/0110002",
    archivePrefix = "arXiv",
    reportNumber = "PU-RCG-01-33",
    doi = "10.1016/S0370-2693(01)01366-1",
    journal = "Phys. Lett. B",
    volume = "524",
    pages = "5--14",
    year = "2002"
}

@article{Moroi:2001ct,
    author = "Moroi, Takeo and Takahashi, Tomo",
    title = "{Effects of cosmological moduli fields on cosmic microwave background}",
    eprint = "hep-ph/0110096",
    archivePrefix = "arXiv",
    reportNumber = "TU-632",
    doi = "10.1016/S0370-2693(01)01295-3",
    journal = "Phys. Lett. B",
    volume = "522",
    pages = "215--221",
    year = "2001",
    note = "[Erratum: Phys.Lett.B 539, 303--303 (2002)]"
}

@article{Baumann:2010tm,
    author = "Baumann, Daniel and Nicolis, Alberto and Senatore, Leonardo and Zaldarriaga, Matias",
    title = "{Cosmological Non-Linearities as an Effective Fluid}",
    eprint = "1004.2488",
    archivePrefix = "arXiv",
    primaryClass = "astro-ph.CO",
    doi = "10.1088/1475-7516/2012/07/051",
    journal = "JCAP",
    volume = "07",
    pages = "051",
    year = "2012"
}

@article{Carrasco:2012cv,
    author = "Carrasco, John Joseph M. and Hertzberg, Mark P. and Senatore, Leonardo",
    title = "{The Effective Field Theory of Cosmological Large Scale Structures}",
    eprint = "1206.2926",
    archivePrefix = "arXiv",
    primaryClass = "astro-ph.CO",
    doi = "10.1007/JHEP09(2012)082",
    journal = "JHEP",
    volume = "09",
    pages = "082",
    year = "2012"
}

@article{Carrasco:2013mua,
    author = "Carrasco, John Joseph M. and Foreman, Simon and Green, Daniel and Senatore, Leonardo",
    title = "{The Effective Field Theory of Large Scale Structures at Two Loops}",
    eprint = "1310.0464",
    archivePrefix = "arXiv",
    primaryClass = "astro-ph.CO",
    doi = "10.1088/1475-7516/2014/07/057",
    journal = "JCAP",
    volume = "07",
    pages = "057",
    year = "2014"
}

@article{Baldauf:2015zga,
    author = "Baldauf, Tobias and Schaan, Emmanuel and Zaldarriaga, Matias",
    title = "{On the reach of perturbative methods for dark matter density fields}",
    eprint = "1507.02255",
    archivePrefix = "arXiv",
    primaryClass = "astro-ph.CO",
    doi = "10.1088/1475-7516/2016/03/007",
    journal = "JCAP",
    volume = "03",
    pages = "007",
    year = "2016"
}

@article{Foreman:2015lca,
    author = "Foreman, Simon and Perrier, Hideki and Senatore, Leonardo",
    title = "{Precision Comparison of the Power Spectrum in the EFTofLSS with Simulations}",
    eprint = "1507.05326",
    archivePrefix = "arXiv",
    primaryClass = "astro-ph.CO",
    doi = "10.1088/1475-7516/2016/05/027",
    journal = "JCAP",
    volume = "05",
    pages = "027",
    year = "2016"
}

@article{Bertolini:2016bmt,
    author = "Bertolini, Daniele and Schutz, Katelin and Solon, Mikhail P. and Zurek, Kathryn M.",
    title = "{The Trispectrum in the Effective Field Theory of Large Scale Structure}",
    eprint = "1604.01770",
    archivePrefix = "arXiv",
    primaryClass = "astro-ph.CO",
    doi = "10.1088/1475-7516/2016/06/052",
    journal = "JCAP",
    volume = "06",
    pages = "052",
    year = "2016"
}

@article{Steele:2021lnz,
    author = "Steele, Theodore and Baldauf, Tobias",
    title = "{Precise Calibration of the One-Loop Trispectrum in the Effective Field Theory of Large Scale Structure}",
    eprint = "2101.10289",
    archivePrefix = "arXiv",
    primaryClass = "astro-ph.CO",
    doi = "10.1103/PhysRevD.103.103518",
    journal = "Phys. Rev. D",
    volume = "103",
    number = "10",
    pages = "103518",
    year = "2021"
}

@article{Chudaykin:2025aux,
    author = "Chudaykin, Anton and Ivanov, Mikhail M. and Philcox, Oliver H. E.",
    title = "{Reanalyzing DESI DR1. I. {\ensuremath{\Lambda}}CDM constraints from the power spectrum and bispectrum}",
    eprint = "2507.13433",
    archivePrefix = "arXiv",
    primaryClass = "astro-ph.CO",
    reportNumber = "MIT-CPT/5890",
    doi = "10.1103/qsnt-dppc",
    journal = "Phys. Rev. D",
    volume = "113",
    number = "6",
    pages = "063502",
    year = "2026"
}

@article{DESI:2024hhd,
    author = "Adame, A. G. and others",
    collaboration = "DESI",
    title = "{DESI 2024 VII: cosmological constraints from the full-shape modeling of clustering measurements}",
    eprint = "2411.12022",
    archivePrefix = "arXiv",
    primaryClass = "astro-ph.CO",
    reportNumber = "FERMILAB-PUB-24-0854-PPD",
    doi = "10.1088/1475-7516/2025/07/028",
    journal = "JCAP",
    volume = "07",
    pages = "028",
    year = "2025"
}

@article{Chen:2021wdi,
    author = "Chen, Shi-Fan and Vlah, Zvonimir and White, Martin",
    title = "{A new analysis of galaxy 2-point functions in the BOSS survey, including full-shape information and post-reconstruction BAO}",
    eprint = "2110.05530",
    archivePrefix = "arXiv",
    primaryClass = "astro-ph.CO",
    doi = "10.1088/1475-7516/2022/02/008",
    journal = "JCAP",
    volume = "02",
    number = "02",
    pages = "008",
    year = "2022"
}

@article{DAmico:2019fhj,
    author = "D'Amico, Guido and Gleyzes, J{\'e}r{\^o}me and Kokron, Nickolas and Markovic, Katarina and Senatore, Leonardo and Zhang, Pierre and Beutler, Florian and Gil-Mar{\'\i}n, H{\'e}ctor",
    title = "{The Cosmological Analysis of the SDSS/BOSS data from the Effective Field Theory of Large-Scale Structure}",
    eprint = "1909.05271",
    archivePrefix = "arXiv",
    primaryClass = "astro-ph.CO",
    doi = "10.1088/1475-7516/2020/05/005",
    journal = "JCAP",
    volume = "05",
    pages = "005",
    year = "2020"
}

@article{Ivanov:2019pdj,
    author = "Ivanov, Mikhail M. and Simonovi{\'c}, Marko and Zaldarriaga, Matias",
    title = "{Cosmological Parameters from the BOSS Galaxy Power Spectrum}",
    eprint = "1909.05277",
    archivePrefix = "arXiv",
    primaryClass = "astro-ph.CO",
    reportNumber = "INR-TH-2019-016, CERN-TH-2019-132",
    doi = "10.1088/1475-7516/2020/05/042",
    journal = "JCAP",
    volume = "05",
    pages = "042",
    year = "2020"
}

@article{Philcox:2022frc,
    author = "Philcox, Oliver H. E. and Ivanov, Mikhail M. and Cabass, Giovanni and Simonovi{\'c}, Marko and Zaldarriaga, Matias and Nishimichi, Takahiro",
    title = "{Cosmology with the redshift-space galaxy bispectrum monopole at one-loop order}",
    eprint = "2206.02800",
    archivePrefix = "arXiv",
    primaryClass = "astro-ph.CO",
    reportNumber = "CERN-TH-2022-092, YITP-22-60",
    doi = "10.1103/PhysRevD.106.043530",
    journal = "Phys. Rev. D",
    volume = "106",
    number = "4",
    pages = "043530",
    year = "2022"
}

@article{DAmico:2022ukl,
    author = "D'Amico, Guido and Donath, Yaniv and Lewandowski, Matthew and Senatore, Leonardo and Zhang, Pierre",
    title = "{The one-loop bispectrum of galaxies in redshift space from the Effective Field Theory of Large-Scale Structure}",
    eprint = "2211.17130",
    archivePrefix = "arXiv",
    primaryClass = "astro-ph.CO",
    doi = "10.1088/1475-7516/2024/07/041",
    journal = "JCAP",
    volume = "07",
    pages = "041",
    year = "2024"
}

@article{Planck:2019kim,
    author = "Akrami, Y. and others",
    collaboration = "Planck",
    title = "{Planck 2018 results. IX. Constraints on primordial non-Gaussianity}",
    eprint = "1905.05697",
    archivePrefix = "arXiv",
    primaryClass = "astro-ph.CO",
    doi = "10.1051/0004-6361/201935891",
    journal = "Astron. Astrophys.",
    volume = "641",
    pages = "A9",
    year = "2020"
}

@article{Jung:2025nss,
    author = "Jung, Gabriel and Citran, Michele and van Tent, Bartjan and Dumilly, L{\'e}a and Aghanim, Nabila",
    title = "{Constraints on primordial non-Gaussianity from Planck PR4 data}",
    eprint = "2504.00884",
    archivePrefix = "arXiv",
    primaryClass = "astro-ph.CO",
    doi = "10.1051/0004-6361/202555283",
    journal = "Astron. Astrophys.",
    volume = "702",
    pages = "A204",
    year = "2025"
}

@article{Chudaykin:2025vdh,
    author = "Chudaykin, Anton and Ivanov, Mikhail M. and Philcox, Oliver H. E.",
    title = "{Reanalyzing DESI DR1. III. Constraints on inflation from galaxy power spectra and bispectra}",
    eprint = "2512.04266",
    archivePrefix = "arXiv",
    primaryClass = "astro-ph.CO",
    reportNumber = "MIT-CTP/5971",
    doi = "10.1103/fhj3-6q4x",
    journal = "Phys. Rev. D",
    volume = "113",
    number = "6",
    pages = "063552",
    year = "2026"
}

@article{Ivanov:2024hgq,
    author = "Ivanov, Mikhail M. and Cuesta-Lazaro, Carolina and Mishra-Sharma, Siddharth and Obuljen, Andrej and Toomey, Michael W.",
    title = "{Full-shape analysis with simulation-based priors: Constraints on single field inflation from BOSS}",
    eprint = "2402.13310",
    archivePrefix = "arXiv",
    primaryClass = "astro-ph.CO",
    reportNumber = "MIT-CTP/5684",
    doi = "10.1103/PhysRevD.110.063538",
    journal = "Phys. Rev. D",
    volume = "110",
    number = "6",
    pages = "063538",
    year = "2024"
}

@article{DAmico:2022gki,
    author = "D'Amico, Guido and Lewandowski, Matthew and Senatore, Leonardo and Zhang, Pierre",
    title = "{Limits on primordial non-Gaussianities from BOSS galaxy-clustering data}",
    eprint = "2201.11518",
    archivePrefix = "arXiv",
    primaryClass = "astro-ph.CO",
    doi = "10.1103/PhysRevD.111.063514",
    journal = "Phys. Rev. D",
    volume = "111",
    number = "6",
    pages = "063514",
    year = "2025"
}

@article{Cabass:2024wob,
    author = "Cabass, Giovanni and Philcox, Oliver H. E. and Ivanov, Mikhail M. and Akitsu, Kazuyuki and Chen, Shi-Fan and Simonovi{\'c}, Marko and Zaldarriaga, Matias",
    title = "{BOSS constraints on massive particles during inflation: The cosmological collider in action}",
    eprint = "2404.01894",
    archivePrefix = "arXiv",
    primaryClass = "astro-ph.CO",
    reportNumber = "RBI-ThPhys-2024-21, MIT-CTP/5698",
    doi = "10.1103/PhysRevD.111.063510",
    journal = "Phys. Rev. D",
    volume = "111",
    number = "6",
    pages = "063510",
    year = "2025"
}

@article{Berghaus:2019whh,
    author = "Berghaus, Kim V. and Graham, Peter W. and Kaplan, David E.",
    title = "{Minimal Warm Inflation}",
    eprint = "1910.07525",
    archivePrefix = "arXiv",
    primaryClass = "hep-ph",
    doi = "10.1088/1475-7516/2020/03/034",
    journal = "JCAP",
    volume = "03",
    pages = "034",
    year = "2020",
    note = "[Erratum: JCAP 10, E02 (2023)]"
}

@article{Bastero-Gil:2014raa,
    author = "Bastero-Gil, Mar and Berera, Arjun and Moss, Ian G. and Ramos, Rudnei O.",
    title = "{Theory of non-Gaussianity in warm inflation}",
    eprint = "1408.4391",
    archivePrefix = "arXiv",
    primaryClass = "astro-ph.CO",
    doi = "10.1088/1475-7516/2014/12/008",
    journal = "JCAP",
    volume = "12",
    pages = "008",
    year = "2014"
}

@article{Berera:1995wh,
    author = "Berera, Arjun and Fang, Li-Zhi",
    title = "{Thermally induced density perturbations in the inflation era}",
    eprint = "astro-ph/9501024",
    archivePrefix = "arXiv",
    reportNumber = "AZPH-TH-94-8",
    doi = "10.1103/PhysRevLett.74.1912",
    journal = "Phys. Rev. Lett.",
    volume = "74",
    pages = "1912--1915",
    year = "1995"
}

@article{Berera:1996nv,
    author = "Berera, Arjun",
    title = "{Thermal properties of an inflationary universe}",
    eprint = "hep-th/9601134",
    archivePrefix = "arXiv",
    reportNumber = "PSU-TH-167",
    doi = "10.1103/PhysRevD.54.2519",
    journal = "Phys. Rev. D",
    volume = "54",
    pages = "2519--2534",
    year = "1996"
}

@article{Berera:1999ws,
    author = "Berera, Arjun",
    title = "{Warm inflation at arbitrary adiabaticity: A Model, an existence proof for inflationary dynamics in quantum field theory}",
    eprint = "hep-ph/9904409",
    archivePrefix = "arXiv",
    reportNumber = "VAND-TH-99-04",
    doi = "10.1016/S0550-3213(00)00411-9",
    journal = "Nucl. Phys. B",
    volume = "585",
    pages = "666--714",
    year = "2000"
}

@article{LopezNacir:2011kk,
    author = "Lopez Nacir, Diana and Porto, Rafael A. and Senatore, Leonardo and Zaldarriaga, Matias",
    title = "{Dissipative effects in the Effective Field Theory of Inflation}",
    eprint = "1109.4192",
    archivePrefix = "arXiv",
    primaryClass = "hep-th",
    reportNumber = "SLAC-PUB-14995",
    doi = "10.1007/JHEP01(2012)075",
    journal = "JHEP",
    volume = "01",
    pages = "075",
    year = "2012"
}

@article{Bastero-Gil:2011rva,
    author = "Bastero-Gil, Mar and Berera, Arjun and Ramos, Rudnei O.",
    title = "{Shear viscous effects on the primordial power spectrum from warm inflation}",
    eprint = "1106.0701",
    archivePrefix = "arXiv",
    primaryClass = "astro-ph.CO",
    doi = "10.1088/1475-7516/2011/07/030",
    journal = "JCAP",
    volume = "07",
    pages = "030",
    year = "2011"
}

@article{Chaussidon:2024qni,
  author = {Chaussidon, E. and others},
  title = {{Constraining primordial non-Gaussianity with DESI 2024 LRG and QSO samples}},
  eprint = {2411.17623},
  archiveprefix = {arXiv},
  primaryclass = {astro-ph.CO},
  reportnumber = {FERMILAB-PUB-24-0884-PPD},
  doi = {10.1088/1475-7516/2025/06/029},
  journal = {JCAP},
  volume = {06},
  pages = {029},
  year = {2025},
}

@article{Chen:2017ryl,
  author = {Chen, Xingang and Wang, Yi and Xianyu, Zhong-Zhi},
  title = {{Schwinger-Keldysh Diagrammatics for Primordial Perturbations}},
  eprint = {1703.10166},
  archiveprefix = {arXiv},
  primaryclass = {hep-th},
  doi = {10.1088/1475-7516/2017/12/006},
  journal = {JCAP},
  volume = {12},
  number = {12},
  pages = {006},
  year = {2017},
}

@article{Senatore:2009gt,
  author = {Senatore, Leonardo and Smith, Kendrick M. and Zaldarriaga, Matias},
  title = {{Non-Gaussianities in Single Field Inflation and their Optimal Limits from the WMAP 5-year Data}},
  eprint = {0905.3746},
  archiveprefix = {arXiv},
  primaryclass = {astro-ph.CO},
  doi = {10.1088/1475-7516/2010/01/028},
  journal = {JCAP},
  volume = {01},
  number = {1},
  pages = {028},
  year = {2010},
}

@article{2018JCAP...04..030S,
  author = {{Simonovi{\'c}}, Marko and {Baldauf}, Tobias and {Zaldarriaga}, Matias and {Carrasco}, John Joseph and {Kollmeier}, Juna A.},
  title = {{Cosmological perturbation theory using the FFTLog: formalism and connection to QFT loop integrals}},
  journal = {\jcap},
  year = {2018},
  month = apr,
  volume = {2018},
  number = {4},
  eid = {030},
  pages = {030},
  doi = {10.1088/1475-7516/2018/04/030},
  archiveprefix = {arXiv},
  eprint = {1708.08130},
  primaryclass = {astro-ph.CO},
  adsurl = {https://ui.adsabs.harvard.edu/abs/2018JCAP...04..030S},
}

@article{Ivanov:2024xgb,
    author = "Ivanov, Mikhail M. and Obuljen, Andrej and Cuesta-Lazaro, Carolina and Toomey, Michael W.",
    title = "{Full-shape analysis with simulation-based priors: Cosmological parameters and the structure growth anomaly}",
    eprint = "2409.10609",
    archivePrefix = "arXiv",
    primaryClass = "astro-ph.CO",
    reportNumber = "MIT-CTP/5762",
    doi = "10.1103/PhysRevD.111.063548",
    journal = "Phys. Rev. D",
    volume = "111",
    number = "6",
    pages = "063548",
    year = "2025"
}

\end{document}